\documentclass[sigconf]{acmart}
\renewcommand\footnotetextcopyrightpermission[1]{}
\setcopyright{none} 

\usepackage{xspace}
\usepackage{balance}

\usepackage[symbol]{footmisc}
\usepackage{filecontents}
\usepackage{graphicx}
\usepackage{hhline}
\usepackage{xspace}
\usepackage{balance}
\usepackage{ulem}
\usepackage{url}
\usepackage{titlesec}
\usepackage{colortbl}
\usepackage{multirow}
\usepackage{multicol}

 \usepackage{tikz}

\usepackage{amsmath}
\usepackage{enumerate}
\usepackage{enumitem}

\usepackage{algorithmicx}
\usepackage{algorithm}
\usepackage{algpseudocode}
\usepackage{varwidth}

\usepackage{amsthm}
\usepackage{textcomp}
\usepackage{gensymb}
\usepackage{makecell}
\usepackage{tabularx, booktabs}
\usepackage{array}
\usepackage{times}
\usepackage{fullpage}
\usepackage{dsfont}
\usepackage{epsfig}
\usepackage{microtype}
\usepackage{subcaption}
\usepackage{mathtools}
\usepackage{cleveref}
\usepackage{todonotes}
\usepackage{bm}
\usepackage{amsthm}

\newcommand{\numfaults}{\ensuremath{f}\xspace}
\newcommand{\numprocesses}{\ensuremath{n}\xspace}

\newcommand{\initiate}{\textsc{initiate}\xspace}
\newcommand{\echo}{\textsc{echo}\xspace}
\newcommand{\ready}{\textsc{ready}\xspace}
\newcommand{\abort}{\textsc{yield}\xspace}

\newcommand{\checkpoint}{\textsc{checkpoint}\xspace}

\newcommand{\true}{\textsc{true}\xspace}
\newcommand{\false}{\textsc{false}\xspace}

\newcommand{\brb}{\textsc{VRB}\xspace}
\newcommand{\brbsmall}{vrb\xspace}

\newlist{VRB}{enumerate}{1}
\setlist[VRB]{label=\textbf{BRB\arabic*:},leftmargin=4\parindent}
\newcommand{\brbcast}{\textsc{VRB-BCAST}}
\newcommand{\brbdeliver}{\textsc{VRB-DELIVER}}

\newcommand{\primitiveprefix}{\textsc{BBCA}\xspace}
\newcommand{\system}{\textsc{BBCA-LEDGER}\xspace}
\newcommand{\primitive}{\textsc{BBCA}\xspace}
\newcommand{\primitiveproperty}{\textsc{BBCA}}
\newcommand{\primitivesmall}{bbca\xspace}
\newlist{primitivelist}{enumerate}{1}
\setlist[primitivelist]{label=\textbf{BBCA\arabic*:},leftmargin=4\parindent}
\newcommand{\bbcacast}{\textsc{BBCA-BCAST}\xspace}
\newcommand{\bbcadeliver}{\textsc{BBCA-DELIVER-COMMIT}\xspace}
\newcommand{\bbcaabort}{\textsc{BBCA-FINALIZE}\xspace}
\newcommand{\bbcaannounce}{\textsc{BBCA-DELIVER-ADOPT}\xspace}
\newcommand{\bbcaadopt}{\textsc{BBCA-DELIVER-ADOPT}\xspace}  
\newcommand{\plsend}{\textsc{PL-SEND}}
\newcommand{\pldeliver}{\textsc{PL-RECV}}

\newcommand{\primitivelog}{\textsc{\primitive-LOG}\xspace}

\newcommand{\logsmall}{log\xspace}
\newcommand{\propose}{\textsc{PROPOSE}\xspace}
\newcommand{\init}{\textsc{INIT}\xspace}
\newcommand{\eagercommit}{\textsc{FINAL}\xspace}
\newcommand{\commit}{\textsc{COMMIT}\xspace}
\newcommand{\finalize}{\textsc{FINALIZE}\xspace}

\newcommand{\vbc}{\textsc{VBC}\xspace}
\newcommand{\bcpropose}{\textsc{VBC-PARTICIPATE}\xspace}
\newcommand{\bcdecide}{\textsc{VBC-DECIDE}\xspace}

\newcommand{\vbtob}{\textsc{VBTOB}\xspace}

\newcommand{\vbtobdecide}{\textsc{VBTOB-DECIDE}\xspace}

\let\emptyset\varnothing

\newcommand{\snmin}{\mathit{sn}_{\mathit{min}}}
\newcommand{\emptyslot}{-1\xspace}

\newtheorem{theorem}{Theorem}[section]
\newtheorem{lemma}[theorem]{Lemma}

\newcommand{\event}[3]{
    \ifthenelse
    {\equal{#3}{}}
    {\ap{#1.\textrm{#2}}}
    {\ap{#1.\textrm{#2} \mid #3}}
}

\algnewcommand\Instance[2]{\State #1, \textbf{instance} #2}
\algnewcommand\InstanceSystem[3]{\State #1, \textbf{instance} #2, \textbf{system} #3}
\algnewcommand\Trigger[3]{\State \textbf{trigger} $\event{#1}{#2}{#3}$}
\algnewcommand\Schedule[1]{\State \textbf{schedule} $#1$}
\algnewcommand\Cancel[1]{\State \textbf{cancel} $#1$}
\algnewcommand\Broadcast[1]{\State \textbf{broadcast} $#1$}
\algnewcommand\Import[1]{\State \textbf{import} $#1$}

\algnewcommand\Not{\textbf{ not }}
\algnewcommand\AndT{\textbf{ and }}
\algnewcommand\OrT{\textbf{ or }}
\algnewcommand\In{\textbf{ in }}

\algsetblockdefx[Implements]{Implements}{EndImplements}{}{}{\textbf{Implements:}}{}
\algsetblockdefx[Uses]{Uses}{EndUses}{}{}{\textbf{Uses:}}{}
\algsetblockdefx[Init]{Init}{EndInit}{}{}{\textbf{Init:}}{}
\algsetblockdefx[Parameters]{Parameters}{EndParameters}{}{}{\textbf{Parameters:}}{}
\algsetblockdefx[Struct]{Struct}{EndStruct}{}{}[1]{\textbf{Struct} $#1$ \textbf{contains}}{}

\algsetblockdefx[Upon]{Upon}{EndUpon}{}{}[1]{\textbf{upon} $#1$ \textbf{do}}{}
\algsetblockdefx[UponExists]{UponExists}{EndUponExists}{}{}[2]{\textbf{upon exists} $#1$ \textbf{such that} $#2$ \textbf{do}}{}
\algsetblockdefx[UponCondition]{UponCondition}{EndUponCondition}{}{}[1]{\textbf{upon} $#1$ \textbf{do}}{}
\algsetblockdefx[UponReceiving]{UponReceiving}{EndUponReceiving}{}{}[1]{\textbf{upon receiving} $#1$ \textbf{do}}{}
\algsetblockdefx[UponEvent]{UponEvent}{EndUponEvent}{}{}[1]{\textbf{upon event} $#1$ \textbf{do}}{}

\algsetblockdefx[Process]{Process}{EndProcess}{}{}[2]{\textbf{process} \emph{#1}($#2$) \textbf{:}}{}
\algsetblockdefx[Function]{Function}{EndFunction}{}{}[2]{\textbf{function} \emph{#1}($#2$) \textbf{:}}{}
\algsetblockdefx[Procedure]{Procedure}{EndProcedure}{}{}[2]{\textbf{procedure} \emph{#1}($#2$) \textbf{is}}{}
\algsetblockdefx[FunctionNoArgs]{FunctionNoArgs}{EndFunctionNoArgs}{}{}[1]{\textbf{function} \emph{#1}() \textbf{:}}{}

\algdef{SE}[DoUntil]{DoUntil}{EndDoUntil}{\textbf{do}}[1]{\textbf{until} $#1$}
\algsetblockdefx[Struct]{Struct}{EndStruct}{}{}[1]{\textbf{Struct} $#1$ \textbf{contains}}{}
\algsetblockdefx[IfExists]{IfExists}{EndIfExists}{}{}[2]{\textbf{if exists} $#1$ \textbf{such that} $#2$ \textbf{then}}{\textbf{end if}}
\algsetblockdefx[While]{While}{EndWhile}{}{}[1]{\textbf{while} $#1$ \textbf{do}}{\textbf{end while}}
\algsetblockdefx[NTimes]{NTimes}{EndNTimes}{}{}[1]{\textbf{for} $#1$ \textbf{times do}}{\textbf{end for}}
\algsetblockdefx[For]{For}{EndFor}{}{}[3]{\textbf{for} $#1 \in #2..#3$ \textbf{do}}{\textbf{end for}}
\algsetblockdefx[ForAll]{ForAll}{EndForAll}{}{}[2]{\textbf{for} $#1 \in #2$ \textbf{do}}{\textbf{end for}}

\graphicspath{{figures/}}

\begin{document}
\normalem
\fancyhf{} 
\fancyhead{}
\fancyfoot[C]{\thepage}

\title{\textbf{\system:\\ High Throughput Consensus meets Low Latency}}

\author{Chrysoula Stathakopoulou}
\affiliation{%
  \institution{Chainlink Labs}
  \country{}
}

\author{Michael Wei}
\affiliation{%
  \institution{Chainlink Labs and VMWare Research}
  \country{}
}

\author{Maofan Yin}
\affiliation{%
  \institution{Chainlink Labs}
  \country{}
}

\author{Hongbo Zhang}
\affiliation{%
  \institution{Chainlink Labs and Cornell University}
  \country{}
}

\author{Dahlia Malkhi}
\affiliation{%
  \institution{Chainlink Labs}
  \country{}
}

\date{\today}

\begin{abstract}
This paper presents \system%
\footnote{\primitiveprefix, pronounced ``bab-ka'', loosely stands for Byzantine Broadcast with Commit-Adopt.},
a Byzantine log replication technology for partially synchronous networks enabling blocks to be broadcast in parallel, 
such that each broadcast is \emph{finalized} independently and instantaneously into an individual slot in the log.
Every finalized broadcast is eventually \emph{committed} to the total ordering, 
so that all network bandwidth has utility in disseminating blocks.
Finalizing log slots in parallel achieves both high throughput and low latency.

\system is composed of two principal protocols that interweave together, a low-latency/high-throughput happy path, and a high-throughput DAG-based fallback path.

The happy path employs a novel primitive called \primitive,
a consistent broadcast enforcing unique slot numbering.
In steady state, \primitive ensures that a transaction can be committed with low latency, in just 3 network steps.

Under network partitions or faults, 
we harness recent advances in BFT and build 
a fallback mechanism on a direct acyclic graph (DAG) created by \primitive broadcasts. 
In this manner, \system exhibits the throughput benefits of DAG-based BFT in face of gaps. 
\end{abstract}

\maketitle

\section{Introduction}
\label{sec:introduction}

The popularity of blockchains has undeniably changed the landscape of Byzantine fault tolerant (BFT) consensus.
Scalable protocols with multiple leaders, enabling transaction proposals to occur in parallel, are being pursued both in theory and in practice
~\cite{miller2016honey,crain2021red,yang2022dispersedledger, RCC, Mir, ISS, lev2019fairledger, 
 baird2016swirlds, danezis2018blockmania, gkagol2019aleph, danezis2022narwhal, spiegelman2022bullshark, keidar2022cordial, fino}. 
While these protocols demonstrate terrific throughput, they all have a similar ``recipe'': 
They allow parallel broadcasts to proceed for a while, and then wait for a leader to drive a commit decision on an entire ``wave'' of accumulated broadcasts. 
This introduces a rather large delay, as these protocols first wait for reliable broadcasts to be finalized and then start ordering them.

Traditional, non-BFT consensus protocols have benefited greatly from decoupling finalization and sequencing. 
Distributed shared logs~\cite{balakrishnan2012corfu, balakrishnan2013tango, wei2017vcorfu, balakrishnan2020virtual, 10.1145/3477132.3483544}
have shown that finalization and sequencing can be performed independently, and can greatly improve parallelism, throughput and latency. 
Instead of waiting for a consensus protocol to both finalize transactions and order them into a sequenced log, 
they assign slots in the log for transactions by a ``sequencer'' immediately upon request.
Sequenced transactions are then finalized into their assigned slots independently and in parallel, enabling the protocol to proceed as
quickly as the sequencer can assign slots.

In this work, \system, we apply the benefits of decoupling finalization from sequencing to the Byzantine setting. 
How do we build a sequencer in the Byzantine setting, though? Our first insight is remarkably simple:
consistent broadcast itself prevents sender equivocation, 
hence broadcasts can be finalized into pre-assigned slots so long as the uniqueness of slot numbers is enforced by the protocol.
Therefore, as in non-BFT settings, broadcasts carrying unique slot numbers can be finalized independently and in parallel, 
successfully growing a sequenced log.
Based on this insight, \system is built on a core broadcast primitive called \primitive, 
a novel variant of consistent broadcast guaranteeing unique slot numbering.
When there is no contention over slots numbers, \primitive broadcasts can be finalized in three network delays.

However, gaps in the log may occur in \system due to faulty nodes or network partitions. 
We use a fallback consensus decision to finalize gap slots. 
Harnessing recent advances in Byzantine fault tolerance on DAGs~\cite{spiegelman2022bullshark,fino},
we allow broadcasts to keep being finalized and drive consensus decisions with zero message overhead.
In this manner, while a gap in slot numbers may delay growing the committed log, 
all broadcasts have utility in contributing to throughput and system resources continue being fully utilized.

Combining these two components enables \system to decouple finalization and sequencing, providing a optimistically fast, high-throughput happy path with a high-thoughput fallback.
More specifically, \primitive allows blocks of transactions to be finalized into a sequenced log extremely fast in the common case. 
At the same time, using a DAG to drive consensus allows \system to maintain the high-throughput scalability of state-of-art DAG protocols (e.g.,~\cite{gkagol2019aleph, danezis2022narwhal, spiegelman2022bullshark}) under contention.
Note that this line of research work emphasizes separating transaction dissemination from block finalization to improve throughput. 
This is orthogonal yet compatible with the core design of \system.

As an added benefit, \system supports ``eager'' notifications which allow nodes to observe finalized slots in the log out of order.
Many use-cases can benefit from the eager finalization notifications, so that \system can support diverse applications on the same platform.

This paper presents a complete \system protocol, which is part of a larger ongoing system project. The goal of this paper is to
present the background on the key insights of \system, to formally describe the protocol and the key primitive \primitive,
and to argue correctness. We are in the process of building a complete system around \system~which
leverages its full performance and sequencing capabilities.
\section{Model}
\label{sec:perliminaries}
%
We assume a set of \numprocesses nodes, such that each one of them represents an independent trust authority.
A computationally bound adversary controls the network and may corrupt up to \numfaults nodes, such that $\numprocesses \ge 3\numfaults + 1$.
We refer to the corrupted nodes as Byzantine, as they may fail arbitrarily, and to the non-corrupted nodes as correct. 

Each node participates in various types of roles/activities, encapsulated by processes hosted on the same node:
\begin{itemize}
	\item \textit{Validation}: Validator processes are responsible for populating the log with blocks. Each node hosts a single validator process.
	\item \textit{Work}: Worker processes are responsible for transaction dissemination. Each node hosts at least one worker process.
	\item \textit{Ticketing}: A ticketmaster process is a functionality responsible for assigning log slots to validators (see \Cref{sec:ticketmaster} for dertails). A ticketmaster may be implemented by process hosted by nodes, in which case each node hosts at most one ticketmaster process. 
\end{itemize}
Below, we will interchangably refer to a node and to the processes it hosts, since there is a 1-1 mapping between them.
Furthermore, we assume that the adversary controls any process of a corrupted node.

Unless mentioned otherwise, processes communicate through authenticated point-to-point channels.
We assume an eventually synchronous network~\cite{DworkLS88}
such that the communication between any pair of correct processes is asynchronous before a global stabilization time ($GST$), unknown to the processes, 
when the communication becomes synchronous.
Finally, we assume a public key infrastructure (PKI) such that each process $i$ is equipped with a public-private key pair $PK_i$,$SK_i$ used for authentication.
We use $\langle \mathit{[field]}\ldots, \sigma_i\rangle$ to denote messages carrying a signature $\sigma_i$ by $i$.
\section{System Overview}
\label{sec:overview}
\system is a Byzantine log replication technology enabling blocks to be broadcast in parallel, such that each broadcast block is \emph{finalized} independently and instantaneously into an individual slot in the log.
Every finalized block is eventually \emph{committed} to the total ordering, 
so that all network bandwidth has utility in disseminating blocks.
Finalizing log slots in parallel addresses our goal of having both high throughput and low latency.

\subsection{The Log Abstraction}
\label{sec:log-abstraction}
\system exposes an abstraction of a totally ordered log to the set of validators. 
The log is a sequence of indexable \emph{slots} that contain \emph{blocks} or
a special $\bot$ value.
To add a block to the log, \system interface exposes to validators a \propose event call.
We say that a slot is \emph{finalized} once it is filled with a finalized block or a $\bot$.
Slots become finalized consistently and irreversibly across all correct validators.

\begin{figure}[h]
  \centering
  \includegraphics[width=\columnwidth]{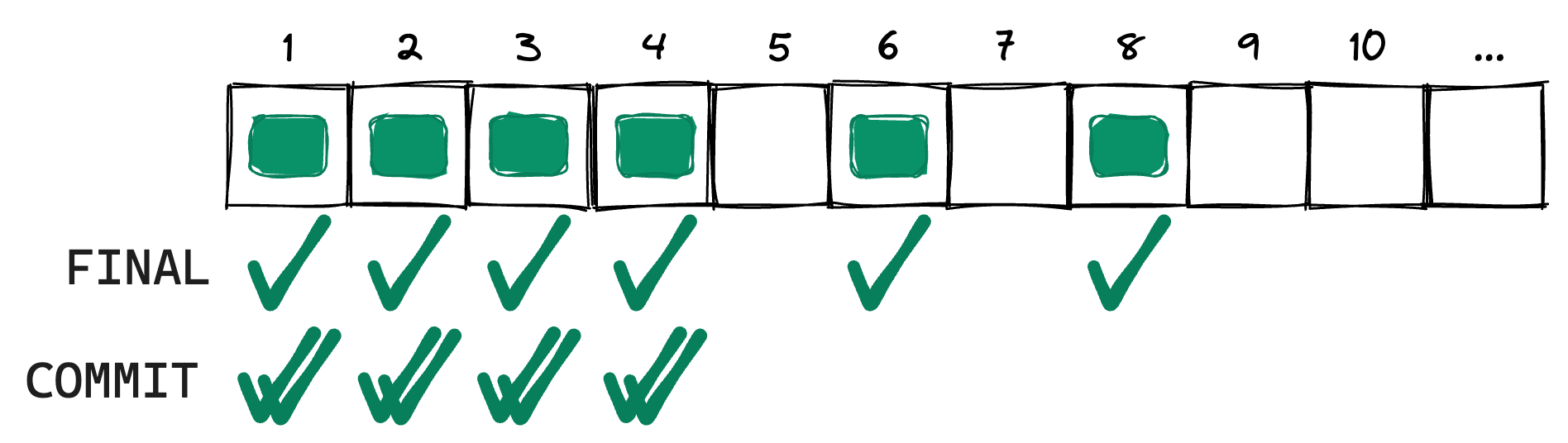}
  \includegraphics[width=\columnwidth]{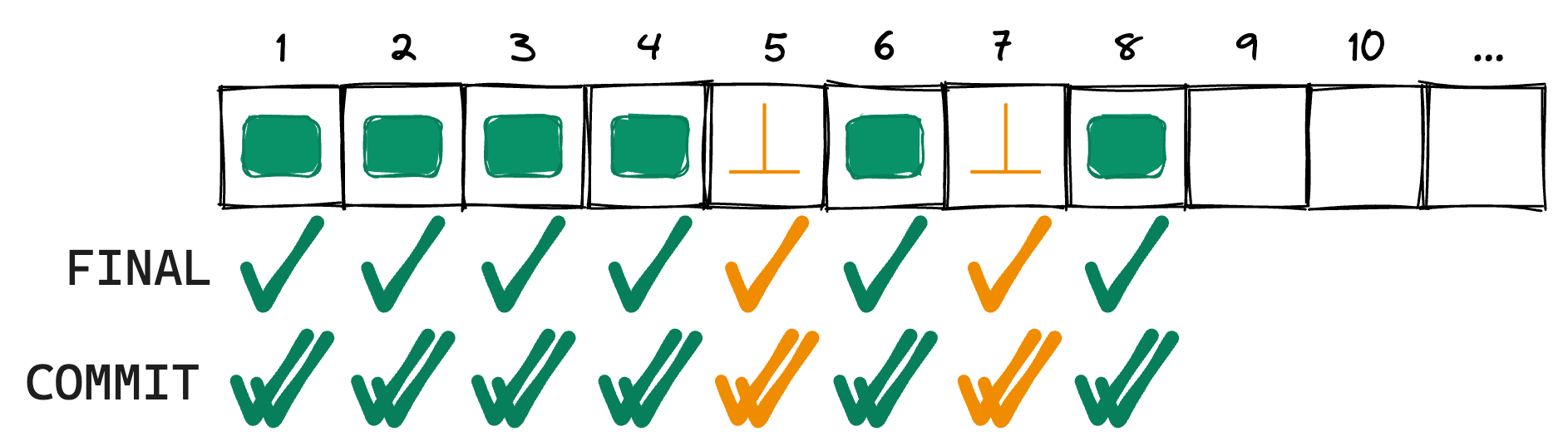}
  \caption{\system. The top shows slots $1-4$ are finalized contiguously, and a \commit notification is sent. 
Slots $6$ and $8$ are finalized but their \commit is pending until $5$ and $7$ become finalized.
On the bottom, slots $5$ and $7$ become permanent holes via a fallback consensus decision, allowing a commit notification to be sent on slots $6$ and $8$.}
  \label{fig:log}
\end{figure}

In common settings, 
validators are notified through a \commit event when a contiguous prefix of the log, starting with the genesis, containing finalized slots grows.
\Cref{fig:log}(top) shows a potential log instance with slots $1-4$ finalized and committed. 

In case ``holes'' are left in the log, they are finalized with a special $\bot$ value via a fallback consensus mechanism, allowing higher slots in the log to become committed. 
\Cref{fig:log}(bottom) shows pending slots ($5$ and $7$) becoming finalized with permanent holes, allowing slots $6$ and $8$ to become committed.

\subsection{Technical Approach}

\system is composed of two principal protocols that interweave together, a low-latency/high-throughput happy path, and a high-throughput DAG-based fallback path.

\paragraph{Happy Path.}
In order to allow blocks to be added to the log in parallel,
we introduce a broadcast primitive called \primitive, a variant of Byzantine consistent broadcast~\cite{CachinGR11} that with two features: 
First, processing broadcasts includes enforcing unique slot numbers. 
Second, broadcasts expose two API events, \bbcaadopt and \bbcadeliver, with a commit-adopt guarantee: \bbcadeliver occurs only it \bbcaadopt occurs at $2f+1$ validators.

Blocks of transactions are broadcast in parallel, as soon as validators are ready to add them to the log.
Every validator may invoke \primitive to add a block to the log with a \emph{desired} slot number.
To use \primitive,
a validator prepares a block with some payload, a set of \emph{causal} references to a causal-frontier of previously delivered \primitive blocks, 
a field to identify its sender, and metadata.  
Multiple validators may propose blocks for different slot numbers simultaneously.
Strategies for slot allocation will be discussed later.
A correct validator finalizes a block $b$ when it consistently delivers $b$ via \primitive
and it has finalized all the blocks in $b$'s causal references.

In the common case, 
when \primitive is invoked by a correct sender with a unique slot number $sn$ and with block $b$, all correct validators eventually receive
 a \commit event for $b$ at slot $sn$.

\paragraph{Fallback Path.}

We use a fallback consensus protocol to finalize slots that remain empty due to faults.
To address the issue of empty slots, 
we build a fallback mechanism on a direct acyclic graph (DAG) created by \primitive broadcasts. 
In this manner, while a gap in the log may delay the \commit event for higher slots, 
the log continues to be filled so the resources of the system are continuously utilized. 
Moreover, consensus decisions about empty slots are embedded within normal broadcasts and have throughput utility as well.
The fallback mechanism works as follows.

Each validator maintains a timer for $sn$, the first empty slot it knows of.
Upon timeout, the validator uses a special \finalize\ mechanism 
that causes the validator to stop participating in \primitive for slot $sn$, 
and to broadcast a ``complaint''.
The complaint carries an intermediate \bbcaadopt event, which \primitive exposes, 
that has the following guarantee:
if \bbcadeliver of a block $b$ with slot $sn$ ever occurs at any correct validator, 
then $f+1$ correct validators will incur \bbcaadopt with $b$ upon invoking \finalize.
We then use a fallback consensus protocol to finalize $sn$ with either a special $\bot$ value, 
or with the (unique) block $b$ indicated in complaints as \finalize-adopted values.

The specific fallback consensus protocol in \system is implemented on top of a DAG using the in-DAG consensus framework introduced in 
the partial-synchrony part of Bullshark~\cite{spiegelman2022bullshark} and generalized to general DAGs in Fino~\cite{fino}.
Briefly, utilizing Fino, 
a \emph{leader-proposal} which causally follows \finalize\ complaints from a quorum of validators determines a value, possibly $\bot$, for the stalled slot number.
A proposal becomes committed through $f+1$ successor blocks interpreted as \emph{votes}.

\paragraph{Ticketmaster.}
Key to the performant functioning of \system, and ultimately to its progress,
is a contention-free allocation of slot numbers to validators. 
We capture the abstract functionality of slot numbering by a role referred to as a ticketmaster.
A naïve ticketmaster strategy is a simple round robin rotation among all validators.
Even such a naïve regime has certain desirable features, including fairness and guaranteed liveness.
However, \system would get delayed by the slowest sender.
More elaborate regimes are discussed in \Cref{sec:ticketmaster}, embodying greater adaptability and dynamism.

\subsection{Enhancements}

\system includes two principal enhancements over the basic protocol, slotless-blocks and \eagercommit events.

\paragraph{Slotless blocks.}

\begin{figure}[h]
  \centering
  \includegraphics[width=\columnwidth]{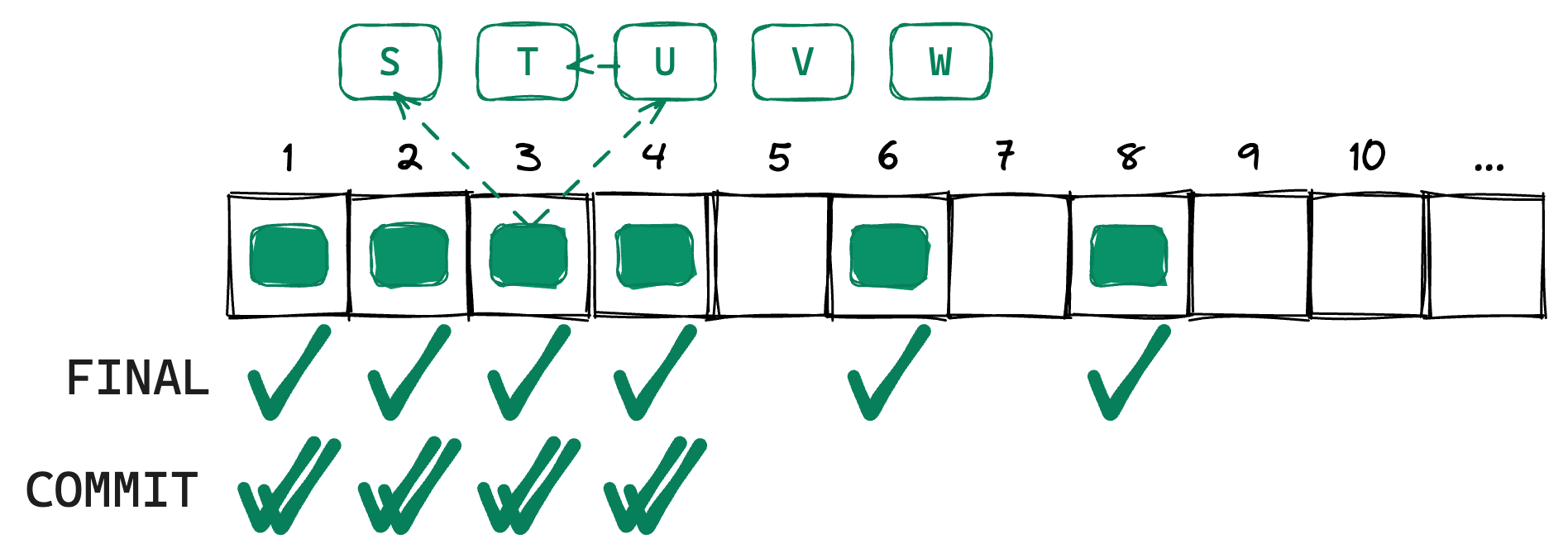}
  \caption{Slotless blocks $S$, $T$, $U$, $V$, $W$. Slot 3 causal predecessors, $S$, $T$, and $U$, become finalized and committed jointly with it.
\eagercommit notifications are received on slots $6$ and $8$.}
  \label{fig:slotless}
\end{figure}

In case of contention over slot numbers, 
at most one invocation of \primitive with the same $sn$ will succeed.
\primitive salvages the blocks which could not be delivered with a slot number by allowing a failed validator to 
\emph{yield} the slot $sn$ and be delivered as a \emph{slotless} block.
The yield mechanism ensures \primitive liveness, and also ensures
that the bandwidth used for dissemination of yielded-blocks is not wasted.
This is similar to uncle blocks~\cite{sompolinsky2015secure}, effectively flattening a sub-graph of blocks linked to a single slot number.
After \primitive delivers a block $b$ with an empty slot number,
another block $b'$ may reference $b$ as a causal predecessor. 
When $b'$ becomes committed, all its causal predecessors become committed as well 
(using any deterministic ordering among them). 
In \Cref{fig:slotless}, blocks $S$, $T$, $U$, $V$, $W$ are slotless; the block at slot 3 has three causal predecessors, $S$, $T$, and $U$, that become committed jointly with it.

\paragraph{Eager finality.}

\system supports an \eagercommit event in addition to \commit events, such that validators are notified as soon as individual slots become finalized, possibly out-of-order.
\Cref{fig:slotless} shows 
slots $6$ and $8$ already finalized and a \eagercommit notification is sent on the blocks in these slots.
Exposing finalized slots \emph{eagerly} via \eagercommit 
notifications is optional. Using \commit offers a standard, totally-ordered log replication service to validators,
whereas \eagercommit may be used to support new use-cases.
By mixing both interfaces, it is possible to construct applications that utilize total ordering for consensus operations (such as abort, or reconfiguration) while employing \eagercommit for steady state. We describe several possible use-cases of \eagercommit.

\begin{figure}[h]
  \centering
  \includegraphics[width=\columnwidth]{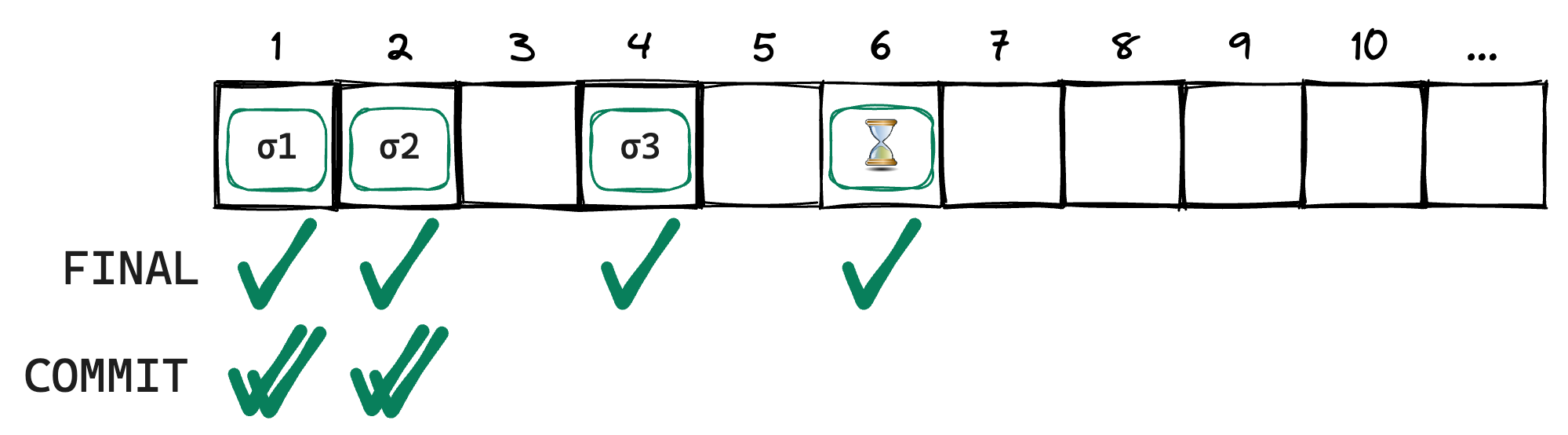}
  \includegraphics[width=\columnwidth]{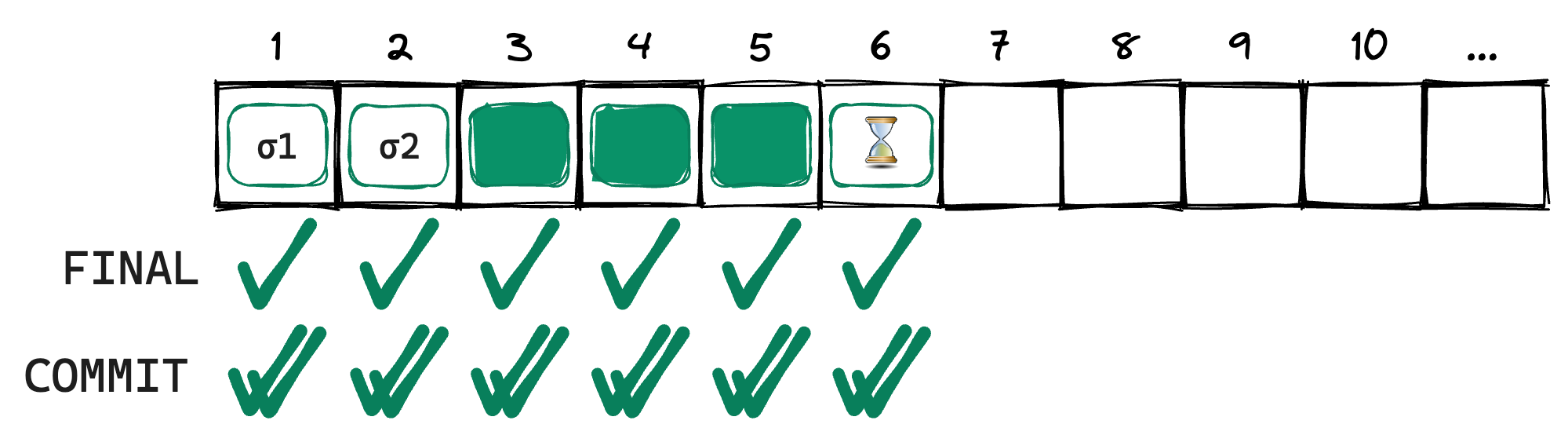}
  \caption{A $3$-out-of-$4$ threshold signature with a deadline example.
  On the top, signature shares $\sigma1$,$\sigma2$, and $\sigma3$ can be aggregated as soon as they become finalized. 
On the bottom, the block which indicates a deadline is committed before $3$ signature shares are committed, and, therefore, the signature reconstruction is aborted.}
  \label{fig:threshold}
\end{figure}

As observed previously in~\cite{baudet2020fastpay, rocket2019scalable, guerraoui2019consensus}, supporting payments requires only partial ordering, that can be readily supported via \eagercommit events. 
Some blockchains, e.g., Sui~\cite{sui} and Linera~\cite{linera}, support single-owner tokenized assets requiring only FIFO ordering of an owner's operations. They provide cross-owner ordering for compound transactions by explicit request.
\system can support owner-only ordering by allowing owners to subscribe to \eagercommit events for their own operations. 
Cross-owner ordering is automatically facilitated by waiting for \commit events in sequence order.

Another example is a $t$-out-of-$n$ threshold signature reconstruction with a deadline.
In this example, the signature can be reconstructed as soon as $t$ signature shares are finalized.
The total order is only required to enforce the deadline, i.e. abort the reconstruction if fewer than $t$ shares are added to the log by a certain slot. 
\Cref{fig:threshold} demonstrates a $3$-out-of-$4$ example.
\section{The \primitive primitive}
\label{sec:bbca}

A core building block of \system is \primitive, 
a variant of Byzantine consistent broadcast~\cite{CachinGR11}
with \emph{desired} slot numbers and an external validity predicate $\mathcal{P}$.
The key property we require from \primitive is that slot numbers of delivered blocks are unique across \primitive instances.
Notice that, uniqueness of slot numbers could be expressed as an external validity predicate; 
we make it an explicit part of the primitive specification in order to express properties about the log which is constructed via these slots.

A \primitive instance for some block $b$ is invoked with an
$\bbcacast(id, sn, b, \mathcal{P})$ event, where $id\in\mathbb{N}$ is the instance label
and $sn\in\mathbb{N}$ is the desired slot number.
We say in short that the sender \primitivesmall-bcasts $(id, sn, b)$.
Two distinct instances invoked with the same desired slot number $sn$ are \emph{conflicting}, 
and at most one can succeed in delivering a block with $sn$.

In order to handle faults that potentially leave slots unfilled,
\primitive allows validators to trigger finalization of slot numbers via a \finalize mechanism that triggers an intermediate \bbcaadopt event that has the following guarantee:
If any validator \bbcadeliver's a block $b$ with slot $sn$, 
then $f+1$ correct validators would incur \bbcaadopt of $b$ upon invoking \finalize.
Adopting is supported through a $\bbcaabort(sn)$ interface.
To use the \finalize mechanism, when $\bbcaabort(sn)$ is invoked by a validator,
\primitive stops participating in broadcasts for slot $sn$, and
asynchronously responds to the validator with an adopt event
$\bbcaannounce(id, sn, b, \pi, \mathcal{Q})$ with a (unique) block $b$ (possibly $\bot$). 
$\pi$ is a certificate of uniqueness, and $\mathcal{Q}$ is a predicate indicating whether the block indicated by $\pi$ should be finalized.
We introduce this API event call to allow any \system slot to become finalized within a finite time, potentially with a $\bot$ value.

In order to salvage a \primitive that cannot obtain its desired slot number and uphold broadcast Validity, 
we allow blocks to be delivered without slots, indicated with a $\emptyslot$ slot value.
Delivery events are denoted $\bbcadeliver(id, sn, b)$, where $sn\in\mathbb{N}\cup\{\emptyslot\}$.
We say that the validator \primitivesmall-delivers $(id, sn, b)$.

\begin{figure}[h]
  \centering
  \includegraphics[width=\columnwidth]{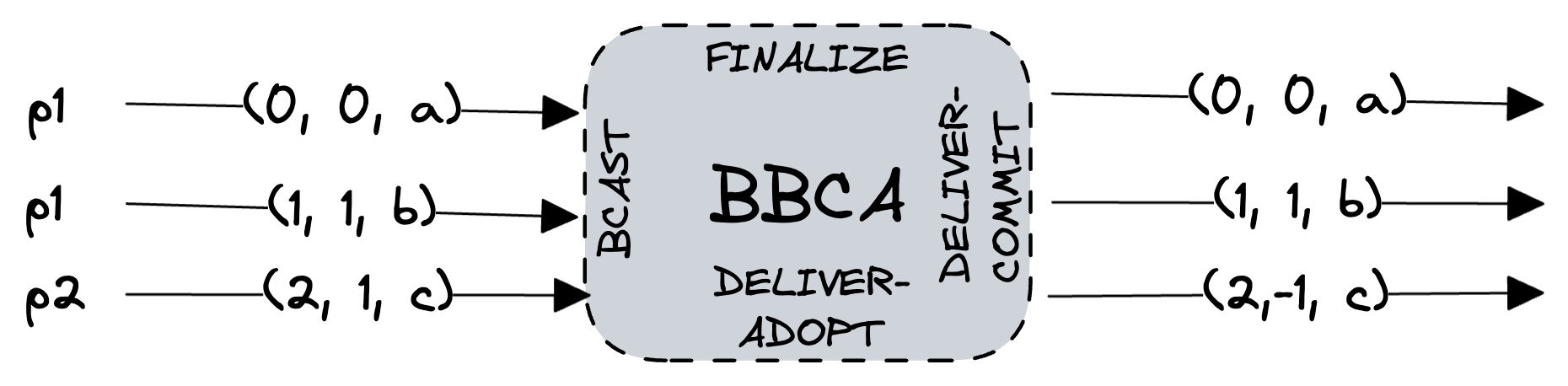}
  \caption{\primitive invocations. Block $c$ is delivered without a slot number, as at most one block can be delivered with slot number $1$.}
  \label{fig:bbca}
\end{figure}

\Cref{fig:bbca} shows an example with multiple \primitive invocations.

We define \primitive with the following properties:

\begin{primitivelist}
	\item \textit{Integrity}: If a correct validator \primitivesmall-delivers $(id, sn, b)$ and the sender $p$ is correct, then $p$ had previously invoked \primitivesmall-bcast for $(id, sn, b)$.
	\item \textit{No-duplication}: Every correct validator \primitivesmall-delivers at most once per instance.
    \item \textit{Consistency}: If a correct validator $p$ \primitivesmall-delivers $(\bm{id}, sn, b)$ and a correct $q$ validator \primitivesmall-delivers $(\bm{id}, sn', b')$  for the same instance $id$, then $b=b'$ and $sn=sn'$ .
    \item \textit{Unique sequencing}: If a correct validator $p$ \primitivesmall-delivers $(id, \bm{sn}, b)$ and a correct validator $q$  \primitivesmall-delivers $(id', \bm{sn}, b')$ for the same non-nil slot number $sn \in \mathbb{N}$, then $id=id'$ (and by Consistency also $b=b'$).
	\item \textit{Validity}: If a correct validator $p$ \primitivesmall-bcasts $(id, sn, b)$ then eventually every correct validator \primitivesmall-delivers $(id, sn', b)$, such that either $sn'=sn \in\mathbb{N}$ or $sn'=\emptyslot$.
	\item \textit{Weak Sequencing Validity}: After GST, if a correct validator $p$ \primitivesmall-bcasts $(id, sn, b)$ with $sn\in\mathbb{N}$, no other validator \primitivesmall-bcasts $(id',sn, b')$,
	and no correct validator incurs $\bbcaabort(sn)$, 
	then every correct validator delivers $(id, sn, b)$.
	\item \textit{External Validity:} If some correct validator \primitivesmall-delivers $(id, sn, b)$ then $\mathcal{P}(b)$ is \true.
	\item \textit{Finalization:} If some correct validator $p$ incurs\\ $\bbcaabort(sn)$ with $sn \in\mathbb{N}$, then $p$ incurs $\bbcaannounce$.
        Moreover, if $2f+1$ validators with indexes in set $I$ incur \\
        $\bbcaannounce(id_i, sn, b_i, \pi_i, \mathcal{Q})$ such that $\mathcal{Q}(\bigcup_{i\in I}\{\pi_i\})=\false$, then no correct node \primitivesmall-delivers $(id, sn, b)$ for any $id$ and block $b$.
\end{primitivelist}

In \Cref{alg:bbca-pt1} we describe a protocol that implements \primitive.
The protocol makes use of a point-to-point communication primitive, $\plsend(p,b)$ and $\pldeliver(p,b)$,
corresponding to a process sending a block $b$ to process $p$ and to a process receiving a block $b$ from process $p$ respectively.
Likewise, it makes use of a validated Byzantine reliable broadcast primitive,
$\brbcast(b,\mathcal{Q})$ and $\brbdeliver(p,b,\mathcal{Q})$, where
a block $b$ with a dedicated sender $p$ is broadcast with an external validity predicate $\mathcal{Q}$ (see~\cite{CachinKPS01} for definition).
We assume that all events in \Cref{alg:bbca-pt1} are handled atomically.

The core of the \primitive protocol borrows from Bracha's known scheme~\cite{BrachaT85}.
However, the key difference in our protocol is that if validators cannot deliver a block $b$ with its desired slot number $sn$ within a time period, they must 
allow the sender to broadcast $b$ without a slot number.
To this end, they ``convince'' an honest sender that it is safe to rebroadcast $b$ with $sn=\emptyslot$.
Conversely,
if it is possible that some correct validator has already delivered $b$ with $sn\neq\emptyslot$, a correct sender will convince all correct validators to deliver the same way.

\paragraph{The protocol.}
A dedicated \primitive sender $s$ for instance $id$ sends to all validators an $\langle\initiate, id, sn, b, \sigma_s\rangle$ message.
Upon receiving an \initiate message for slot number $sn$ and block $b$ and upon the validity predicate $\mathcal{P}$ for $b$ becomes true, a correct validator $p$ verifies that it is sent by the dedicated instance sender, and that the signature on it is valid.
Moreover, $p$ verifies that it has not echoed another message for either the same instance $id$ or $sn$.
Then $p$ sends a message $\langle\echo, id, sn, b, \sigma_p\rangle$ to all validators and starts a timer $T_{\mathit{id}}$ for instance $id$.

Upon receiving $2f+1$ matching valid \echo messages from different validators, $q$ sends a $\langle\ready, id, sn, b\rangle$ message to all validators.
After receiving $2f+1$ \ready messages,
$q$ \primitivesmall-delivers $id,sn,b$ and cancels the timer.

\begin{figure}[h]
  \centering
  \begin{subfigure}{\linewidth}
	  \includegraphics[width=\columnwidth]{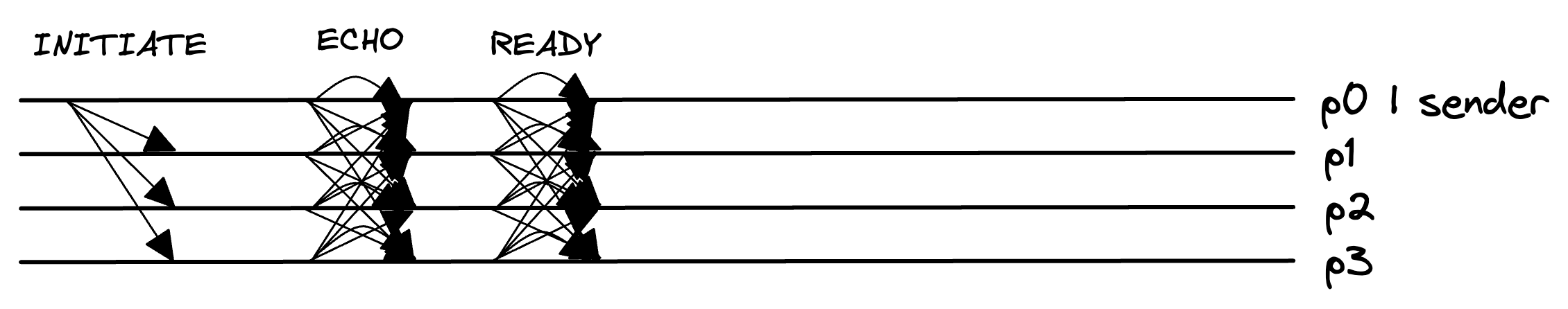}  
 	 \caption{Common case execution.}
 	 \label{fig:bbca-common}
  \end{subfigure}
  \begin{subfigure}{\linewidth}
	  \includegraphics[width=\columnwidth]{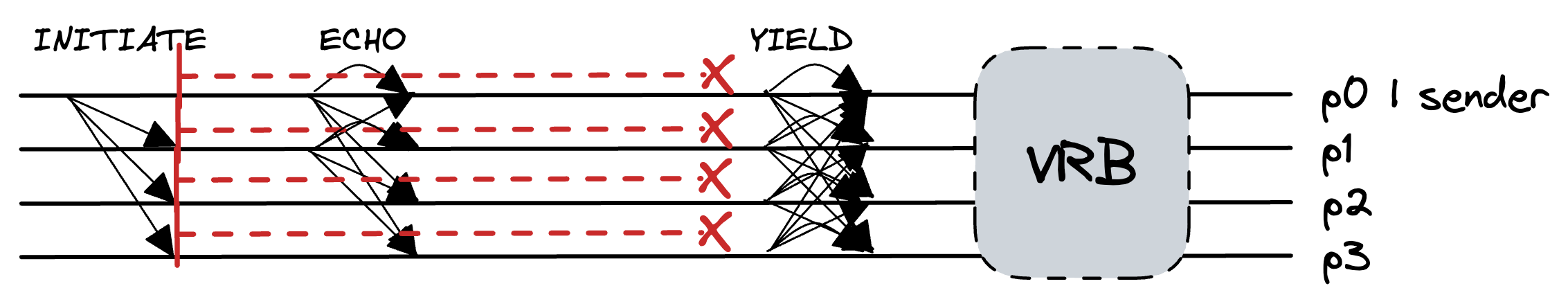}  
 	 \caption{Yield where no node sent a \ready message: validator $0$ must invoke \brb with $\emptyslot$.}
 	 \label{fig:bbca-fault-1}
  \end{subfigure}
   \begin{subfigure}{\linewidth}
	  \includegraphics[width=\columnwidth]{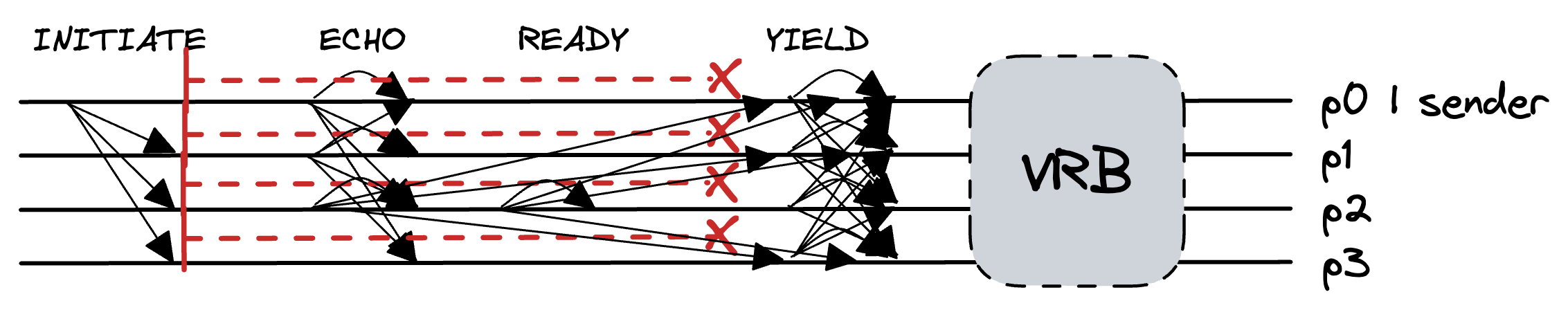}  
 	 \caption{Yield where validator $2$ sent \ready with $sn$: validator $0$ must invoke \brb with $sn$.}
 	 \label{fig:bbca-fault-2}
  \end{subfigure}
  \caption{\primitive executions with and without a timeout for $4$ validators. The dashed line represents the timer.}
  \label{fig:bbca-execution}
\end{figure}

\Cref{fig:bbca-execution}(a) illustrates the common-case execution of our \primitive implementation.

\paragraph{Finalizing a slot number.}
Validators must eventually finalize each slot, potentially with an empty block, 
in order to allow higher slots to commit. 
To this end, 
when the timer of a slot number $sn$ goes off at a correct validator $p$ it invokes $\bbcaabort(sn)$.
After $\bbcaabort(sn)$ is invoked at $p$, $p$ stops accepting and sending \initiate, \echo and \ready messages for $sn$,
and triggers a $\bbcaannounce(sn, b, \pi, \mathcal{Q})$ event, where $\pi=C_R$,
a set of the $2f+1$ \echo messages $p$ has received in case it has sent a \ready message for slot number $sn$,
or an empty set otherwise.
Notice that there exists at most one instance for which $C_R$ could be non-empty.
Contrary to yield, finalizing a slot number requires a consensus decision, which is described in the next section. 

\paragraph{Yielding the slot number of a \primitive instance.}
If the timer of an instance $id$ goes off, $q$ must ``yield'' the slot number of the instance.
To this effect $q$ stops accepting or sending \initiate/\echo/\ready messages for this instance and sends $\langle\abort, id, sn, b, C_R, \sigma_p \rangle$ to all validators.
$C_R$ is a set of the $2f+1$ \echo messages $q$ has received that triggered \ready,
$C_R = \emptyset$ if none.
We refer to $C_R$ as \emph{ready certificate}.
Moreover, upon receiving $f+1$ matching \abort messages from different validators, if $q$ has not already yielded the slot number of the instance, it does so, too.

When a sender of instance $id$ receives $2f+1$ \abort messages for $id$, it must decide how to help all correct validators \bbcadeliver the corresponding block $b$.
If there exists at least one $C_R \neq \emptyset$ with some $sn$ among the \abort messages, it is safe to assign slot number $sn$ to $b$. By quorum intersection, no validator could have received enough \echo messages for a different block $b'$ carrying the same $sn$.
Therefore, the sender invokes $\brbcast((id,sn,\mathcal{C},b),\mathcal{Q})$, where $\mathcal{C}$ is the set of the $2f+1$ ready certificates and the external validity predicate $\mathcal{Q}$ verifies that indeed the set $C$ justifies assigning $sn$ to $b$.

If, on the other hand, all $C_R$ sets are empty, no validator can \primitivesmall-deliver $b$, as not enough validators will ever send a \ready message.
Therefore, it is safe for the sender to rebroadcast $b$ with no slot number.  
It invokes a $\brbcast((id,\emptyslot,\mathcal{C},b),\mathcal{Q})$, where the external validity predicate $\mathcal{Q}$ checks that indeed the set $C$ justifies assigning the special slot number $\emptyslot$ to $b$.

\Cref{fig:bbca-execution}(b-c) illustrate potential yielding scenarios.

\paragraph{Validity.}
\primitive is weaker than consensus, since
we only require to guarantee Validity for a correct sender.
Moreover, even for a single, correct sender, \primitive requires a checkpointing mechanism to satisfy the Validity property for all correct validators.
In particular, note that a correct validator either delivers a block for some instance, or times out.
It is, however, possible that fewer than $f+1$ validators time out for some instance and, therefore, the sender never triggers \brb to help them \primitivesmall-deliver some block for this instance.
To ensure that all correct validators \primitivesmall-deliver a block for any instance with a correct sender, a checkpointing mechanism should be in place.
A correct validator $p$ sends a $\langle\checkpoint, id, sn, b, \sigma_p\rangle$ message upon \primitivesmall-deliver.
A validator may \primitivesmall-deliver $(id, sn, b)$ upon receiving $f+1$ \textsc{checkpoint} messages, as at least one of them is from a correct validator.
Importantly, this mechanism does not increase latency in the common case and, moreover, validators can batch \textsc{checkpoint} messages.

\begin{algorithm}
  \caption{\primitive implementation.}
  \label{alg:bbca-pt1}
  \scriptsize

  \label{alg:sb-impl}
  \begin{algorithmic}[1]
    \Implements
      \State \primitive \textsc{\primitivesmall} with external validity predicates $\mathcal{P}$ and $\mathcal{Q}$.
    \EndImplements
    \Uses
      \State Byzantine reliable broadcast, \textsc{\brbsmall}
      \State Point-to-point links, \textsc{pl}
    \EndUses
    
    \Parameters
      \State $\numprocesses$
      \Comment{Number of nodes}
      \State $\numfaults$
      \Comment{Maximum number of faulty nodes}
      \State $N$
      \Comment{Set of all $\numprocesses$ nodes}
      \State $\Delta$ 
      \Comment{Progress timeout period}
      \State $T\gets\{\}$
      \Comment{Map of timers per instance label}
      \State \textit{sentecho}$\gets\{\}$
      \Comment{Map of boolean values per instance label}
      \State \textit{sentready}$\gets\{\}$
      \Comment{Map of boolean values per instance label}
      \State \textit{delivered}$\gets\{\}$
      \Comment{Map of boolean values per instance label} 
      \State \textit{yielded}$\gets\{\}$
      \Comment{Map of boolean values per instance label} 
      \State \textit{echos}$\gets\{\}$
      \Comment{Map of maps of $(sn,b)$ tuples per sender per instance label}
      \State \textit{readys}$\gets\{\}$
      \Comment{Map of maps of $(sn,b)$ tuples per sender per instance label}
      \State \textit{yields}$\gets\{\}$
      \Comment{Map of maps of $(sn,b)$ tuples per sender per instance label}
      \State \textit{checkpoints}$\gets\{\}$
      \Comment{Map of maps of $(sn,b)$ tuples per sender per instance label}
      \State \textit{readyCertPerId}$\gets\{\}$
      \Comment{Map ready certificates per sender per instance label}
      \State \textit{readyCertPerSn}$\gets\{\}$
      \Comment{Map ready certificates per sender per sequence number}
      \State \textit{seqnos}$\gets\{\}$
      \Comment{Map of boolean values per sequence number}
      \State \textit{finlalized}$\gets\{\}$
      \Comment{Map of boolean values per sequence number}
    \EndParameters

	\UponEvent{\bbcacast(id, sn, b, \mathcal{P})}
	\Comment{Triggered only by the dedicated sender $s$ of instance with label $id$}
		\If{\textit{finalized}$[sn]=\false$}
		\State $\sigma_s\gets\operatorname{sign}(s,(id, sn, b))$
		\ForAll{q}{N}
			\State\textbf{trigger} $\plsend(q, \langle \initiate, id, sn, b, \sigma_s \rangle)$
		\EndForAll
		\EndIf
	\EndUponEvent

	\UponEvent{\pldeliver(q, \langle \initiate, id, sn, b, \sigma_q\rangle)  \textbf{ and } \mathcal{P}(b)=\true}
	\Comment{Executed by node $p$}
		\If{$q$ is the sender of instance with label $id$}
			\State $T[id].\operatorname{settimer}(\Delta)$
        		\Comment{Set a progress timer}
			\If{\textit{sentecho}$[id]=\false$ \textbf{and} \textit{yielded}$[id]=\false$ \textbf{and} \textit{finalized}$[sn]=\false$ \textbf{and} \textit{seqnos}$[sn]=\false$\textbf{and} $\operatorname{valid}(q,(id, sn, b), \sigma_q)$}
				\label{ln:integrity}
				\State\textit{sentecho}$[id]\gets\true$
				\State\textit{seqnos}$[sn]\gets\true$
				
        		\State $\sigma_p\gets\operatorname{sign}(p,(id, sn, b))$
 				\ForAll{q}{N}
					\State\textbf{trigger} $\plsend(q,\langle \echo, id, sn, b, \sigma_p\rangle)$
				\EndForAll
			\EndIf
		\EndIf
	\EndUponEvent
	
	\UponEvent{\pldeliver(q, \langle \echo, id, sn, b, \sigma_q\rangle)}
	\Comment{Executed by node $p$}
		\If{\textit{yielded}$[id] = \false$  \textbf{and} \textit{finalized}$[sn]=\false$ \textbf{and} $\operatorname{valid}(q,(id, sn, b), \sigma_q)$ \textbf{and} \textit{echos}$[id][q]=\bot$}
			\State\textit{echos}$[id][q]\gets (sn,b) $
		\EndIf
	\EndUponEvent	
	
	\UponExists{(id,sn,b)}{|\{p\in N | \operatorname{\textit{echos}}[id][p]=(sn,b)\}|>2f \operatorname{\textbf{and}}  \operatorname{\textit{sentready}[id]=\false }}
	\Comment{Executed by node $p$}
		\State\textit{sentready}$[id]\gets\true$
		\State \textit{readyCertPerId}$\gets\operatorname{\textit{echos}}[id]$
		\State \textit{readyCertPerSn}$\gets\operatorname{\textit{echos}}[id]$
		\State $\sigma_p\gets\operatorname{sign}(p,(id, sn, b))$
		\ForAll{q}{N}
			\State\textbf{trigger} $\plsend(q,\langle \ready, id, sn, b, \sigma_p\rangle)$
		\EndForAll
	\EndUponExists
	
	\UponEvent{\pldeliver(q, \langle \ready, id, sn, b, \sigma_q\rangle)}
	\Comment{Executed by node $p$}
		\If{\textit{yielded}$[id] = \false$  \textbf{and} \textit{finalized}$[sn]=\false$ \textbf{and} $\operatorname{valid}(q,(id, sn, b), \sigma_q)$ \textbf{and} \textit{readys}$[id][q]=\bot$}
			\State\textit{ready}$[id][q]\gets (sn,b) $
		\EndIf
	\EndUponEvent
	
	\UponExists{(id,sn,b)}{|\{p\in N | \operatorname{\textit{readys}}[id][p]=(sn,b)\}|>2f \operatorname{\textbf{and}}  \operatorname{\textit{delivered}[id]=\false }}
	\label{ln:no-duplication-1}
	\Comment{Executed by node $p$}
		\State\textit{delivered}$[id]\gets\true$
		\State $T[id].\operatorname{cancel}()$
		\Comment{Cancel progress timer with with timeout duration $\Delta$}
		\State\textbf{trigger} $\bbcadeliver(p, id, sn, b)$
		\State $\sigma_p\gets\operatorname{sign}(p,(id, sn, b))$
		\ForAll{q}{N}
			\State\textbf{trigger} $\plsend(q,\langle \checkpoint, id, sn, b, \sigma_p\rangle)$ 
		\EndForAll
	\EndUponExists
	
		\algstore{bbca}
  \end{algorithmic}
\end{algorithm}

\begin{algorithm}
    \scriptsize
    \ContinuedFloat
    \begin{algorithmic}[1]
    \algrestore{bbca}
    
    \UponEvent{T[id].\operatorname{timeout}}
	\Comment{Executed by node $p$}
		\State\textit{yielded}$[id]\gets\true$
		\State $C_R\gets\operatorname{\textit{readyCertPerId}}[id]$
		\State $\sigma_p\gets\operatorname{sign}(p,(id, sn, b, C_R))$
		\ForAll{q}{N}
			\State\textbf{trigger} $\plsend(q,\langle \abort, id, sn, b, C_R,\sigma_p\rangle)$ 
		\EndForAll
	\EndUponEvent
	
	\UponEvent{\bbcaabort(sn)}
	\Comment{Executed by node $p$}
		\State\textit{finalized}$[sn]\gets\true$
		\State $C_R\gets\operatorname{\textit{readyCertPerSn}}[sn]$
		\State\textbf{trigger} $\bbcaannounce(p, sn, b, C_R)$ 
		\label{ln:announcement}
	\EndUponEvent
	
	\UponEvent{\pldeliver(q, \langle \abort, id, sn, b, C_R, \sigma_q\rangle)}
	\Comment{Executed by node $p$}
		\If{$\operatorname{valid}(q,(id, sn, b, C_R), \sigma_q)$ \textbf{and} \textit{yields}$[id][q]=\bot$}
			\State\textit{yields}$[id][q]\gets sn, b, C_R $
		\EndIf
	\EndUponEvent
	
	\UponExists{(id,sn,b)}{|\{p\in N | \operatorname{\textit{yields}}[id][p]=(sn,b,\_)\}|>2f \operatorname{\textbf{and}}  \operatorname{\textit{delivered}[id]=\false }}
	\Comment{Executed by sender $s$}
	\State $\mathcal{C}\gets$ \textit{yields}$[sn]$
		\State \textbf{if exists} $(sn,b,C_R)\in\mathcal{C}$  \textbf{such that} $C_R\neq\emptyset$ \textbf{then}
		\State \hskip \algorithmicindent \textbf{trigger} $\brbcast((id, sn, b,\mathcal{C}),\mathcal{Q})$
				\State{else}
		\State \hskip \algorithmicindent \textbf{trigger} $\brbcast((id, \emptyslot, b, \mathcal{C}), \mathcal{Q})$
		\State{end if}
	\EndUponExists
	
	\UponExists{(id,sn,b)}{|\{p\in N | \operatorname{\textit{yields}}[id][p]=(sn,b,\_)\}|>f \operatorname{\textbf{and}}  \operatorname{\textit{yielded}[id]=\false }}
				\State\textit{yielded}$[id]\gets\true$
		\State $C_R\gets\operatorname{\textit{readyCertPerId}}[id]$
		\State $\sigma_p\gets\operatorname{sign}(p,(id, sn, b, C_R))$
		\ForAll{q}{N}
			\State\textbf{trigger} $\plsend(q,\langle \abort, id, sn, b, C_R, \sigma_p\rangle)$ 
		\EndForAll
	\EndUponExists

	\UponEvent{\brbdeliver(p, (id, sn, b), \mathcal{Q})}
		\If{\textit{delivered}[id]=\false }
		 \label{ln:no-duplication-2}
			\State\textit{delivered}$[id]\gets\true$
			\State\textbf{trigger} $\bbcadeliver(p, id, sn, b)$
			\State $\sigma_p\gets\operatorname{sign}(p,(id, sn, b))$
			\ForAll{q}{N}
				\State\textbf{trigger} $\plsend(q,\langle \checkpoint, id, sn, b, \sigma_p\rangle)$ 
			\EndForAll
		\EndIf
	\EndUponEvent
	
	\UponEvent{\pldeliver(q, \langle \checkpoint, id, sn, b, \sigma_q\rangle)}
	\Comment{Executed by node $p$}
	\If{$\operatorname{valid}(q,(id, sn, b), \sigma_q)$ \textbf{and} \textit{checkpoints}$[id][q]=\bot$}
		\State\textit{checkpoints}$[id][q]\gets (sn,b)$
	\EndIf
	\EndUponEvent	
	
	\UponExists{(id,sn,b)}{|\{p\in N | \operatorname{\textit{checkpoints}}[id][p]=(sn,b)\}|>f \operatorname{\textbf{and}}  \operatorname{\textit{delivered}[id]=\false }}
		\State\textit{delivered}$[id]\gets\true$
		\State\textbf{trigger} $\bbcadeliver(p, id, sn, b)$
	\EndUponExists
	
	\caption{\primitive implementation (continues).}	
  	\label{alg:bbca-pt2}
   \end{algorithmic}
\end{algorithm}

\section{The \system}
\label{sec:log}
\system exposes to validators a Byzantine fault-tolerant log abstraction built around the \primitive primitive.
Validators can request to add blocks to the log.
Every validator 
can read from the log, or subscribe to asynchronous notifications when slots are finalized. 

Eventually, every slot is \emph{finalized} with some value.
Values can be blocks, or a special $\bot$ value.
Each slot has two finalization events, \commit and \eagercommit. 

More specifically, we define the log interface as follows:
\begin{itemize}
	\item $\init()$: initializes the log.
	\item $\propose(b)$: proposes to add block $b$.
	\item $\eagercommit(b)$: finalizes $b$.
	\item $\commit(b)$: commits $b$, its non-committed, slotless causal predecessors, and all blocks in its preceding slots in the log.
\end{itemize}

To propose a block $b$, a validator $p$ invokes $\bbcacast(id,sn,b)$ for a new instance $id$ and the next slot $sn$ available for $p$.
Policies for slot allocation to validators are discussed in \Cref{sec:ticketmaster}.

A \emph{block} becomes finalized when all the blocks in its causal references are finalized and either (1) is \primitivesmall-delivered, with or without a slot number, or (2) is decided by the fallback consensus protocol as the value for some slot.
A \emph{slot} $sn$ becomes finalized when a block becomes finalized for $sn$ or a $\bot$ value is decided by the fallback consensus protocol for $sn$. 
The slot becomes committed when the slot is finalized and its previous slot is committed. 
The log grows infinitely and starts with a known committed block (``genesis'').

When a slot becomes finalized at a validator,
\system triggers a \eagercommit notification to the validator. 
The finalized slot value may contain a block, or $\bot$ to mark a ``hole''.
Subsequently, when the slot becomes committed, 
\system triggers a \commit notification.

\paragraph{Fallback finalization protocol.}
Recall that \primitive only guarantees a weak validity property for slot numbers (see \primitiveproperty6 in \Cref{sec:bbca}).
To ensure that any slot is eventually finalized, 
and therefore all blocks added to the log can eventually become committed,
\system employs on a fallback finalization protocol which implements a consensus decision on stalled slots.

A validator $p$ maintains a timer for the first non-finalized slot it knows of as $\snmin$.
Upon finalizing $\snmin$, $p$ restarts a timer for the next non-finalized position.
If the timer expires, $p$ invokes the fallback finalization protocol by triggering the $\bbcaabort$ event for $\snmin$
and subscribes for $\bbcaannounce$ on $\snmin$.

Upon a $\bbcaannounce(sn, b, C_R,\mathcal{Q})$ event, $p$ \primitive's a block which includes in its metadata a $\finalize$ flag and the information $sn, b, C_R$.
If $p$ has not sent a \ready message for $sn$, then $C_R=\emptyset$ and $b=\bot$.

\Cref{alg:bbca-log} describes the log protocol using a consensus protocol with an external validity predicate (\vbc) as a ``blackbox''. 
\vbc exposes two events, $\bcpropose(v,Q)$ is triggered by a validator to participate in the protocol with a value $v$, and $\bcdecide(v)$.
We assume that $\vbc$ satisfies the following properties: if all correct validators participate with some value, eventually, all validators decide a value $v$, which is the value with which some validator participated (liveness) and, no two validators decide differently (safety).
Moreover, any value decided by a correct validator satisfies $\mathcal{Q}$.

Once a validator has \primitivesmall-delivered a set of $2f+1$ 
blocks with a $\finalize$ flag for some sequence number $sn$, it invokes $\bcpropose([sn,b,\mathcal{C}], \mathcal{Q})$, where $\mathcal{C}$ the corresponding set of $2f+1$ ready certificates.
The proposal entails a non-$\bot$ block $b$ if any of the $2f+1$ ready certificate is non-empty.
The external predicate $\mathcal{Q}$ enforces this condition to the value that correct validators deliver via $\vbtob$.
A correct validator finalizes slot $sn$ the first time it receives an event $\bcdecide([sn,b,\mathcal{C}])$ with some block $b$, as soon as it finalizes all $b'$ causal predecessors.

\paragraph{Fallback protocol implementation.}
There exist several consensus protocols in literature which satisfies the properties described above, such as HotStuff~\cite{yin2019hotstuff} and Tendermint~\cite{buchman2018latest}.
We implement the fallback protocol using the consensus framework of Fino~\cite{fino} on top of the DAG which is already populated by including causal references in \primitivesmall-delivered blocks, so that no additional messages are required to be exchanged.
Information pertaining to the consensus protocol is embedded in the metadata field of blocks.
In this way, every \primitive broadcast carries useful payload and allows the system to continue utilizing its bandwidth for transaction dissemination even when resolving non-finalized slots.

Briefly, the Fino consensus framework is structured in views and each view has a designated leader.
Upon entering a new view $r$, the leader broadcasts a proposal for view $r$ with \primitive, by including $proposal(r)$ in the metadata field of the block.
The proposal for view $r$ is justified if it has in its causal predecessors either $f+1$ votes for proposal $r-1$, or $2f+1$ complaints for round $r-1$.
When a proposal is added to the DAG, if it is justified then correct validators vote for it.
To vote for a proposal for view $r$, a validator includes $vote(r)$ in the metadata field of the block.
A consensus proposal is \emph{decided} in the DAG once it has $f+1$ votes succeeding it.
If the proposal for view $r$ is not decided in time, a correct validator \bbcacast\ a complaint by adding $complaint(r)$ in the metadata of the block.
Fino requires that messages by each validator are numbered and authenticated. If a validator delivers some message as the $k^{th}$ from a particular sender, then all other validators deliver the same message as the sender's $k^{th}$ message.
We use the external validity predicate $\mathcal{P}$ of \primitive to enforce this condition.

\begin{figure}[h]
  \centering
  \includegraphics[width=\columnwidth]{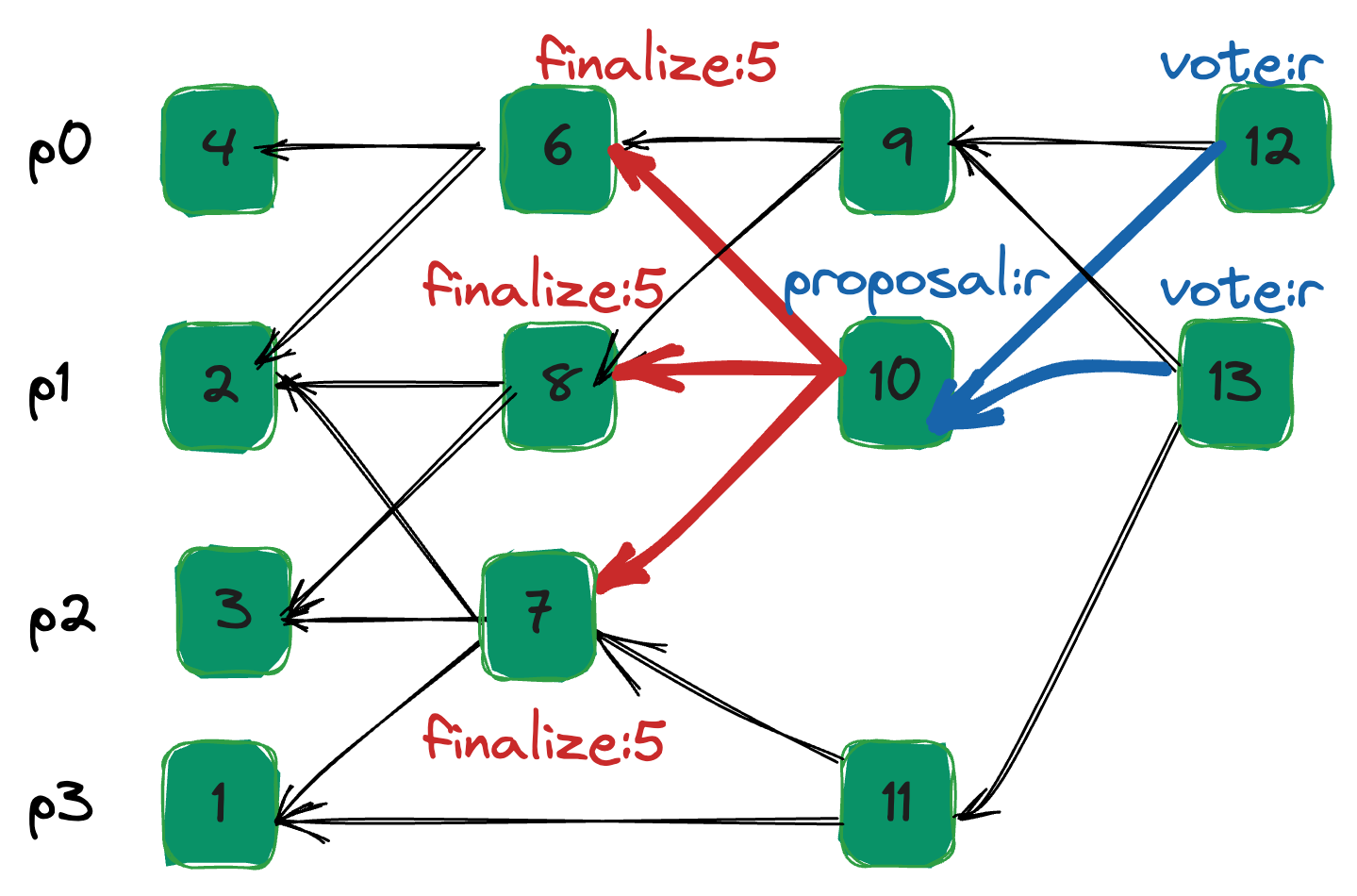}
  \caption{Finalizing slot $4$: Blocks $5$,$7$, and $8$ indicate that validators the timers of $p0$, $p1$, and $p2$ expired for slot $4$.
  	Validator $p1$, who is a consensus leader for view $r$, implicitly proposes block $10$ for slot $4$. Blocks $12$ and $13$ are the votes for the leader proposal. }
  \label{fig:log-fallback}
\end{figure}

The first consensus proposal which follows $2f+1$ predecessor blocks with a $\finalize$ flag for some slot number $sn$, constitutes a consensus proposal to stop waiting for \primitive to finalize the slot $sn$.
It becomes decided by the Fino consensus protocol as described above.
Note that Fino proposals do not need to explicitly carry the set not ready certificates $\mathcal{C}$, as the reference the corresponding blocks which carry them as causal predecessors.
\Cref{fig:log-fallback} shows an example of a DAG instance for finalizing an expired slot number.

\paragraph{Causality.}
When validator $p$ creates a new block, 
it includes in it references to a set of \emph{causal predecessors}, a causal-frontier of previously \primitivesmall-delivered blocks which $p$ has not previously referenced in other blocks.
A reference to block $b$ is simply a pair $(sn,H(b))$ , where $H$ a collision resistant hash function and $sn$ the slot number with which $b$ was delivered, or $\bot$ if $b$ is a slotless block.
A validator before finalizing a block $b$ which it \primitivesmall-delivers, it finalizes all the blocks in 
$b$'s causal predecessors.
In  other words, finalizations are causal.
To ensure causal finalizations, we want to ensure that no correct validator delivers $b$ without delivering $b$'s causal predecessors first.
For this we use the external validity predicate of \primitive.
Moreover, a correct validator, upon receiving a \bbcacast event for a block $b$, starts a timer for all slot numbers of blocks in its $b$'s causal predecessors that are not finalized yet.
If the timer times out and the slot number is still not finalized, $p$ invokes $\bbcaabort(sn)$.
This timeout along with \primitive and the fallback finalization protocol ensure that if a correct validator proposes a block with a block $b'$ in its causal predecessors, 
then all correct validators can eventually \primitivesmall-deliver or decide $b'$ (see \Cref{lemma:totality}).
Therefore, a proposal from a correct validator will always eventually satisfy the external validity predicate.

\paragraph{Slotless blocks.}
Recall that \primitive may deliver slotless blocks without slot numbers.
Slotless blocks are not directly assigned to any slot.
However, validators receive a \eagercommit notification as soon as they become finalized. 
Moreover, when the block $b'$ that references $b$ as a causal predecessor becomes
committed, 
validators receive a \commit notification about $b$ together with $b'$ (using any deterministic order among them). 

\begin{algorithm}
  \caption{\system implementation with validity predicates $\mathcal{P}, \mathcal{Q}$.}
  \label{alg:bbca-log}
  \scriptsize

  \label{alg:sb-impl-pt1}
  \begin{algorithmic}[1]
    \Implements
      \State \primitivelog \textsc{\logsmall}, executed by process $p$.
    \EndImplements
    \Uses
      \State \primitive
      \State Validated Byzantine Total Order Broadcast, \textsc{vbtob}
    \EndUses
    
    \Parameters
    	\State $\numprocesses$
      	\Comment{Number of nodes}
      	\State $\numfaults$
      	\Comment{Maximum number of faulty nodes}
     	\State $N$
     	\Comment{Set of all $\numprocesses$ nodes}
      	\State $\Delta$ 
      	\Comment{Progress timeout period}
		\State $pid\gets 0$
		\Comment{\primitive identifier for node $p$}
		\State \textit{log}
		\Comment{Map of blocks per sot number}
		\State \textit{toFinalize}$\gets\{\}$
      	\Comment{Map of boolean values per slot number}
      	\State \textit{unreferrenced}$\gets\emptyset$
      	\Comment{Set of blocks previously not referenced by some block} 
      	\State \textit{firstUncommitted}$\gets 0$
		\Comment{First uncommitted log slot number}
		\State \textit{finalized}$\gets\emptyset$
		\Comment{Set of finalized blocks}	
		\State \textit{toFinalize}$\gets\emptyset$
		\Comment{Set of (seq no, blocks) tuples pending finalization}	
    \EndParameters
    
    \Struct{Block}
      \State sender\Comment{Validator who proposes the block}
      \State command \Comment{Block metadata}
      \State predecessors \Comment{Block references}
      \State payload \Comment{Transaction references}
    \EndStruct

	\UponEvent{\init()}
		\State $T_f.\operatorname{reset}(\Delta)$
		\Comment{timer with duration $\Delta$ for the first not uncommitted slot}
	\EndUponEvent
	
	\UponEvent{\propose(b)}
		\State $\operatorname{propose}(b,\bot)$
	\EndUponEvent
	
	\UponEvent{\bbcadeliver(\_, sn, b)}
		\State $\operatorname{finalize}(sn, b)$
		\If{$b\notin$\textit{unreferrenced}}
			\State\textit{unreferrenced}$\gets \{b\}\cup$\textit{unreferrenced}
		\EndIf
	\EndUponEvent
	
    \UponEvent{log[\operatorname{\textit{firstUncommitted}}] \neq None}
        \State $b\gets\log[\operatorname{\textit{firstUncommitted}}]$
        \If{$b\neq\bot$}
        	\ForAll{b'}{\operatorname{traverse}(b)}
				\State \textbf{trigger} \commit($b'$)
			\EndForAll
		\EndIf
        \State $T_f.\operatorname{reset}(\Delta)$
        \State \textit{firstUncommitted} $\gets$\textit{firstUncommitted}$+1$
   	\EndUponEvent
	
	\UponEvent{T_f.\operatorname{timeout}}
		\State\textbf{trigger} \bbcaabort(\textit{firstUncommitted})
		\State $T_f.\operatorname{reset}(\Delta)$
		\State \textit{firstUncommitted}$\gets$\textit{firstUncommitted}$+1$
		\While{log[\operatorname{\textit{firstNotTimedOut}}] \neq None}
			\State \textit{firstUncommitted}$\gets$\textit{firstUncommitted}$+1$
		\EndWhile
	\EndUponEvent
	
  \algstore{log}
  \end{algorithmic}
\end{algorithm}

\begin{algorithm}
    \scriptsize
    \ContinuedFloat
    \begin{algorithmic}[1]
    \algrestore{log}
    
	\UponEvent{\bbcaannounce(sn, b, C_R, \mathcal{Q})}
		\State $b'.\operatorname{payload}\gets$ payload()
		\State $\operatorname{propose}(b', [\finalize,sn, b, C_R])$
	\EndUponEvent
	
	\State\textbf{upon event} $\bbcadeliver(\_, \_, b)$ \textbf{such that} $b$.{command}$=[\finalize,sn', b', C_R]$	\textbf{do}
		\State\hskip\algorithmicindent\textbf{if} \textit{toFinalize} $[sn'][q]=\bot$ \textbf{then}
		\State\hskip\algorithmicindent\hskip\algorithmicindent\textit{toFinalize}$[sn'][q]\gets sn', b', C_R $
		\State\hskip \algorithmicindent\textbf{end if}
	\\
	
	\UponExists{(id,sn,b)}{|\{p\in N | \operatorname{\textit{toFinalize}}[sn][p]=(sn,b,C_R)\}|>2f}
		\State $\mathcal{C}\gets$ \textit{toFinalize}$[sn]$
		\State \textbf{if exists} $(sn,m,C_R)\in$ \textit{toFinalize}$[sn]$ \textbf{such that} $C_R\neq\emptyset$ \textbf{then}
		\State \hskip \algorithmicindent \textbf{trigger} $\bcpropose([sn, m,\mathcal{C}], \mathcal{Q})$
		\State{\textbf{else}}
		\State \hskip \algorithmicindent \textbf{trigger} $\bcpropose([sn,\bot,\mathcal{C}], \mathcal{Q})$
		\State{\textbf{end if}}
	\EndUponExists
	
	\UponEvent{\bcdecide(sn, b)}
		\If{$b=\bot$}
			\State \textit{log}$[sn]\gets b$	
			\State \textbf{return}
		\EndIf
		\State $\textit{toFinalize}\gets\{b\}\cup(b.\operatorname{predecessors}\setminus\textit{finalized})\cup\textit{toFinalize}$
	\EndUponEvent
	
	\UponExists{b\in\textit{toFinalize}}{b.\operatorname{predecessors}\in\textit{finalized}}
		\State \textbf{trigger} $\bbcadeliver(\_, sn, b)$
	\EndUponExists

	\UponEvent{\bbcacast(\_, sn, b)}
		\ForAll{b'}{b.\operatorname{predecessors}}
			\If{$b'\notin\textit{finalized}$}
				\State\textbf{trigger} $\operatorname{timeout}(sn,b)$ \textbf{ after } $\Delta$ 
			\EndIf
		\EndForAll
	\EndUponEvent
	
	\UponEvent{\operatorname{timeout}(sn,b)}
			\If{$b\notin\textit{finalized}$}
				\State\textbf{trigger} $\bbcaabort(sn)$ 
			\EndIf
	\EndUponEvent
	
	\Function{propose}{b, c}
		\State $id\gets p||pid$
		\State $pid\gets pid+1$
		\State $sn\gets$ next()
		\State $R\gets$ unreferrenced
		\State unreferrenced$\gets\emptyset$
		\State $b.\operatorname{sender}, b.\operatorname{command}, b.\operatorname{predecessors}
				\gets p, c, R$
		\State \textbf{trigger} \bbcacast($id, sn, b, \mathcal{P}$)
	\EndFunction
	
	\FunctionNoArgs{payload}{}
		\State returns the next batch of transaction references, possibly empty
	\EndFunctionNoArgs
	
	\FunctionNoArgs{next}{}
		\State returns the next $sn$ which $p$ is assigned
	\EndFunctionNoArgs
	 
	\Function{traverse}{p}
		\State traverses a graph with root $p$ and returns a list of its nodes in a deterministic order
	\EndFunction
	
	\Function{finalize}{sn, b}
		\State $\textit{finalized}\gets\{b\}\cup\textit{finalized}$
		\If{$sn\neq\emptyslot$}
			\State \textit{log}$[sn]\gets b$	
			\State \textbf{trigger} \eagercommit($b'$)
		\EndIf
	\EndFunction
	\caption{\system implementation (continues).}	
	\label{alg:sb-impl-pt2}
  \end{algorithmic}
\end{algorithm}

\section{Practical Considerations}

\subsection{Ticketmaster}
\label{sec:ticketmaster}

Key to the good performance of system is an effective way for allocating sequence slots to validators exclusively.
Otherwise, sender broadcasts might ``collide'' in their attempted sequence slots.

It is important to note that in case of collisions, broadcasts are not lost:
\primitive eventually delivers every block \primitivesmall-bcast by a correct sender, albeit without a sequence number. Therefore, eventually a consensus decision will reference it and it will become committed to the total order.  
However, this scenario will suffer the same commit delay as previous protocols, which rely on a leader to commit a wave of accumulated broadcasts; we want to avoid this in \system.
Additionally, eventually blocks carrying sequence numbers need to be enabled for the system to make progress.

To this end, we introduce an abstract functionality referred to as a \emph{ticketmaster}.
The goal of a ticketmaster is to assign sequence slots to actively brodcasting validators, hereon referred to as active validators, exclusively and consecutively.
In order to prevent Byzantine validators from capriciously attempting to \primitivesmall-bcast sequence numbers,
potentially causing collisions and slowdowns,
validators should be allowed to invoke \bbcacast\ only with their sequence slots. 
Validators accept and echo other validators' blocks only when they carry allowed sequence numbers.

There are a number of strategies for realizing a ticketmaster functionality.

\paragraph{Round robin.}
A communication-less strategy is to assign sequence numbers by a round robin rotation among active validators.
For example, say that there are four validators, $p_0$, $p_1$, $p_2$, $p_3$.
By round robin rotation, they will be assigned sequence slots, $0, 1, 2, 3$ respectively, then $4, 5, 6, 7$, and so on.

The advantage of round robin is that it removes contention completely, and that it incurs no extra messages.
The downside is that it requires all validators to be active in each rotation. 
Additionally, any validator that fails to use their slot slows down 
those that come after it because they have to wait for the fallback protocol to finalize the slot.

As a mitigation measure,
the set of active validators may be automatically adjusted every time a fallback consensus decision is made in order to finalize a timed-out slot.
The consensus decision pertaining a new configuration of active validators would have to specify the slot from which the rotation starts.
A new rotation might incur temporary contention due to the switchover, but this does not hurt consistency; it only (potentially) causes a temporary hickup in throughput.

For example, say slot $3$ and $7$ fail to finalize within a designated timeout period.
A fallback consensus can finalize the slots with a $\bot$ value,
and at the same time, will kick validator $p_3$ from the active validators' set
starting at slot number $8$.

To bring $p_3$ back to the system, there would be another fallback consensus decision with a special control command for adding $p_3$ back.

\paragraph{Ticketmaster process.}
Another option is using a ticketmaster server to hand out slot reservations. This works as follows.
When a validator wishes to propose a block, it first reaches out to the ticketmaster process with a ticket-request to be assigned a slot. The ticketmaste responds with a certified slot, which the validator can use to invoke \bbcacast with.
Note that the slot must be certified by the server to prevent Byzantine validators from capriciousy attempting to \bbcacast\ sequence numbers.
Due to the chance of Byzantine faults,
the ticketmaster process should be rotated regularly among nodes every fixed number of slots.

The advantage of this approach is there only active validators need to broadcast.
The downside is that the interaction with a ticketmaster process adds two network latencies to the critical path of \primitive.
However, note that the messages involved in this interaction are very small.
Additionally, batching can amortize the communication cost and delay across a bulk of transmissions.

Experience gained in the benign fault setting~\cite{balakrishnan2012corfu, balakrishnan2020virtual}
indicates that this strategy can perform extremely well,
easily reaching hundreds of thousands to a million commits per second.
The ticketmaster process, albeit being a bottleneck, is very lightweight and stateless, and therefore can perform
with very high throughput.

\paragraph{Hybrid.}

In a hybrid approach, slots are allocated by a rotational regime.
However, each rotation (or group of rotations) is determined by a ticketmaster process on demand.
This works as follows.

Except for the first rotation, each validator sends a ticket request to the ticketmaster process of the next rotation.
The ticketmaster process waits for a certain threshold of requests (a threshold may be determined by assistant policy),
and then \primitivesmall-bcasts a message allocating validators to the next set of rotations.

An advantage of the threshold participation requirement
is guaranteeing a certain degree of participation fairness, namely,
that some fraction of the slots in each rotation will be assigned to correct validators.
This is similar to the full rotation regime, however, the hybrid regime
has the benefit of adaptively determining a rotation by demand.
The downside is the same as having a ticketmaster per request, namely
having a server in the critical path of broadcast deliveries.

\subsection{Transaction dissemination}
Similarly to Narwhal~\cite{Narwhal}, we decouple transaction dissemination from establishing a total order.
Transactions can be disseminated by a consistent broadcast protocol run by worker processes.
Each validator may correspond to multiple worker processes.
Transaction deduplication techniques, as in \cite{Mir} or \cite{RCC} can be applied among workers to ensure effective bandwidth utilisation.
We do not go in depth with respect to transaction dissemination, as this is still a work in progress.

Also, a practical implementation of BBCA would avoid repeatedly sending the actual user payload in \echo, \ready, and \checkpoint messages, but instead with block digests whenever possible.
One can pull the full payload from the validator that first sends a block and a correct validator only needs to do it once throughout the entire \primitive instance. Therefore, in our implementation,
an auxiliary $\langle\textsc{pull}\rangle$ message is used to trigger the resend of an \echo/\ready/\checkpoint message with the full payload in place of the hash. An adversary cannot effectively abuse this pull-style functionality as
a validator needs to resend a full payload at most once to each of other correct validators. It does not impose additional latency when every validators just receive the payload in the \initiate message and also handles the scenario
where the asynchrony of the network could make an \echo/\ready message arrive earlier than \initiate message, with the minimum communication overhead.  Another optimization would be to precautiously send an \abort message upon receiving an \initiate message with a conflicting sequence number.
Last, we follow PBFT~\cite{PBFT} which requires signed messages. However, we could replace signatures with authenticated point-to-point channels, similarly to~\cite{PBFT:2002}.

\subsection{Pacing dissemination via the DAG}
\primitive broadcasts form a DAG, whose primary purpose is the fallback consensus mechanism. 
However,
We can utilize the DAG itself to prevent the use of the DAG for fallback. What we mean by this is the following.

Optimistically, \system should operate at its full speed
even without ever reverting to the  fallback consensus. In theory, one can create contentionless and gap-less
broadcasts.
We are therefore designing \system to synchronize the pace of 
block generations among validators. This is needed because validators would otherwise
keep working on their own slots, where the fastest stays way ahead of others,
leaving lots of gaps which don't contribute to the commit progress. While a
process-based ticketmaster could balance the pace, it is more natural to just
make use of the DAG itself for synchronization. It also helps unify the happy path with
the fallback path as there is already some causality among the blocks. Our current implementation
uses a simple strategy: each validator can at most have $k$ outstanding \primitive instances.
Whenever some instances are finalized, the validator can immediately dispatch new broadcasts. A validator
keeps track of the frontier set (blocks without causal successors), which is used as the references
for a generated block. The frontier is then set to only contain the new block
once it is finalized. This design does not require extra timeouts and
relies on the causality of the DAG to help synchronize the pace.

\section{Correctness Arguments}
\label{sec:guarantees}
\subsection{\primitive implementation}
\label{sec:log-correctness}
In this section we argue that \primitive implementation in \Cref{alg:bbca-pt1} satisfies \primitive properties.
\subsubsection{Integrity}
\label{sec:bbca-integrity}
Integrity is satisfied 
by correct validators only sending a \echo message for an \initiate message signed by the dedicate sender of the instance
(see \Cref{alg:bbca-pt1}, line \ref{ln:integrity}).

\subsubsection{No duplication}
\label{sec:bbca-no-duplication}
No duplication trivial to satisfy by a correct validator checking that it has not \primitivesmall-delivered for the same instance before (see \Cref{alg:bbca-pt1}, lines \ref{ln:no-duplication-1}, \ref{ln:no-duplication-2}).
\subsubsection{Consistency}
\label{sec:bbca-consistency}
Let's assume by contradiction that there exist correct validators $p$ and $q$ who \primitivesmall-deliver $(id, sn, b)$ and $id, sn', b'$ with $sn\neq sn'$ or $b\neq b'$.

There are three conditions which, upon being satisfied, cause a validator to trigger $\bbcadeliver(p, id, sn, b)$: (a) by receiving $2f+1$ matching \ready messages before a timeout for instance $id$ or a $\bbcaabort(sn)$ event, (b) because of a $\brbdeliver$ event, and (c) by receiving $f+1$ matching \checkpoint messages. 

We exhaustively examine all the possible combinations.
\begin{enumerate}
	\item Both $p$ and $q$ trigger \bbcadeliver\ by condition (a).
	\item Both $p$ and $q$ trigger \bbcadeliver\ by condition (b).
	\item One of $p$ and $q$ triggers \bbcadeliver\ by condition (a) and one of $p$ and $q$ triggers \bbcadeliver\ by condition (b).
	\item Both $p$ and $q$ trigger \bbcadeliver\ by condition (c).
	\item One of $p$ and $q$ triggers \bbcadeliver\ by conditions (a) or (b) and one of $p$ and $q$ triggers \bbcadeliver\ by condition (c).
	\item One of $p$ and $q$ triggers \bbcadeliver\ before a timeout on instance $id$ or \bbcaabort\ for $sn$ or $sn'$ has been triggered and the other after.
\end{enumerate}

\paragraph{Case 1.} In this case, $p$ has received $2f+1$ \echo messages for $(id, sn, b)$ and $q$ has received $2f+1$ \echo messages for $(id, sn', b')$, which suggests that there exists a correct validator who sent and \echo message for both triplets.
A contradiction, since a correct validator sends only one \echo message per instance (see \Cref{alg:bbca-pt1}, line \ref{ln:integrity}).

\paragraph{Case 2.}
A contradiction due to the consistency property of \brb (no two different correct validators trigger \brbdeliver for different messages) .

\paragraph{Case 3.}
Let's assume, without loss of generality, that $p$ \primitivesmall-delivers before a timeout on instance $id$ or a \bbcaabort\ event (condition (a))  and $q$ after (condition (b)).
Therefore $p$ has received $2f+1$ \echo messages for $(id, sn, b)$.
This suggests that the external validity predicate for \brb is satisfied: there exist some ready certificate $C_R\neq\emptyset$ for $id, sn', b'$, i.e., $2f+1$ validators sent an \echo message for $id, sn', b'$.
A contradiction, since a correct validator sends only one \echo message per instance.

\paragraph{Case 4.}
In this case, $p$ has received $f+1$ \checkpoint messages from $f+1$ validators for $(id, sn, b)$, among which there exists at least one correct validator, let's call it $p1$.
Similarly, $q$ has received $f+1$ \checkpoint messages from $f+1$ validators for $(id, sn', b')$, among which there exists at least one correct validator, let's call it $q1$.
If $p1=q1$, then $p1$ has triggered \bbcadeliver for different triplets, a contradiction.
If $p1\neq p2$, the have triggered \bbcadeliver due to either (a) or (b). Hence, we fall back to cases $1,2$, and $3$.

\paragraph{Case 5.}
Let's assume, without loss of generality, that $p$ has received $f+1$ \checkpoint messages from $f+1$ validators for $(id, sn, b)$,  among which there exists at least one correct validator, let's call it $p1$.
In this case $p1$ has triggered \bbcadeliver due to either (a) or (b).
Therefore, we fall back to cases $1,2$, and $3$ for the correct validators $p1$ and $q$.

\subsubsection{Unique sequencing}
\label{sec:bbca-sequencing}
Again, let's assume by contradiction that the there exist correct validators $p$ and $q$ who \primitivesmall deliver $(id, sn, b)$ and $id', sn, b'$ with $id\neq id'$ and $b\neq b'$.
Since $sn\neq\emptyslot$, both $p$ and $q$  trigger \bbcadeliver\ before a timeout on instance $id$ or $id$ and before \bbcaabort\ for $sn$ .

Therefore, $p$ has received $2f+1$ \echo messages for $(id, sn, b)$ and $q$ has  has received $2f+1$ \echo messages for $(id', sn, b')$, which suggests that there exists a correct validator who sent and \echo message for both triplets.
A contradiction, since a correct validator sends only one \echo message per slot number (see \Cref{alg:bbca-pt1}, line \ref{ln:integrity}).

\subsubsection{Validity}
\label{sec:bbca-validity}
\paragraph{Synchronous network communication.} If the network is synchronous, i.e. we are after GST, no correct validator triggers a \bbcaabort\ event and there is no invocation of \bbcacast\ with the same slot number $s$, by the weak sequencing validity property, every correct validator delivers the same triplet $(id,sn,b)$.

Let's assume now that we are before GST.

\paragraph{Timed-out instances.} 
There exist two mutually exclusive cases: (a) at least $f+1$ correct validators who timed-out for instance $id$ and (b) at most $f$ validators timed-out for instance $id$.

In case (a), all correct validators will receive a \abort message and, therefore, send an \abort message themselves.
Therefore, the sender of the instance will receive $2f+1$ \abort messages and invoke $\brbcast(id, sn', b)$ where  $sn'=sn$ or $sn'=\emptyslot$, depending on the ready certificates in the \abort messages.
Since the sender is correct, by \brb validity, all correct validators will \brb-deliver $(id,sn',b)$, and, therefore, all correct validators who had timed-out before sending an \abort message \primitivesmall-deliver $id, sn', b$.
Finally, by \primitive consistency, all correct validators will \primitivesmall-deliver $id, sn', b$.

In case (b), there exist at least $f+1$ correct validators who \primitivesmall-deliver $(id,sn,b)$.
They all deliver $(id,sn,b)$ by \primitive consistency property. 
Therefore, all correct validators, who had not delivered before, will eventually receive $f+1$ \checkpoint messages and \primitivesmall-deliver $(id,sn,b)$.

\paragraph{Conflicting instances.} Let us now assume that some other sender invokes $\bbcacast(id',sn,b')$.
If some correct validator sends an \echo message for $(id',sn',b')$ before receiving an \initiate message for $(id,sn,b)$, then 
it is possible that some correct validators will not gather $2f+1$ \echo messages for $(id,sn,b)$, resulting in a timeout for instance $id$. 
Then validity follows by the same arguments as in timed-out instances paragraph.

\paragraph{Finalizations.} If some correct validator triggers a \bbcaabort\ event before it \primitivesmall-delivers, it will timeout for instance $id$.
 Then validity follows by the same arguments as in timed-out instances paragraph.

\subsubsection{Weak sequencing validity}
\label{sec:bbca-weak-seq-validity}
Since there is no other instance \primitivesmall-bcast  with $sn$, no correct validator has sent an $\langle \echo, id', sn, b'\rangle$ message.
Also, since the sender is correct, it invokes \bbcacast for $id,sn,b$ such that $\mathcal{P}(id,sn,b)=\true$. 
Moreover, since we are after GST no correct validator triggers $\bbcaabort(sn)$ and no timeout occurs for $id$.
Therefore, every correct validator sends an $\langle \echo, id, sn, b\rangle$ message and, thus, every correct validator receives $2f+1$  valid \echo messages and sends a \ready message.
Similarly, every correct validator receives $2f+1$ valid \ready messages and \primitivesmall-delivers $(id,sn,b)$ 
Finally, by consistency and unique sequencing properties, all validators \primitivesmall-deliver $(id,sn,b)$.

\subsubsection{External validity}
\label{sec:consistency}
If $P(b)$ is not satisfied, no correct validator sends an \echo message.
Therefore, no correct validator can deliver $(id,sn,b)$.

\subsubsection{Finalization}
\label{sec:bbca-finalization}
A $\bbcaabort(sn)$ event trivially triggers a \bbcaannounce\ event by the implementation (see \Cref{alg:bbca-pt1}, line \ref{ln:announcement}).
Let's now argue about the second part of the property.
In our implementation the proof $\pi$ is a ready certificate.
Predicate $\mathcal{Q}$ corresponds to no ready certificate being empty.
Therefore, we have $2f+1$ empty ready certificates, among which at least $f+1$ are from correct validators.
Since $f+1$ correct validators did not sent a \ready message, there exist at most $2f$ validators which sent a \ready message for slot number $sn$.
Therefore, no correct validator could have \primitivesmall-delivered $id, sn, b$ for any block $b$ and any instance $id$.

\subsection{\system implementation correctness}
\label{sec:log-correctness}

Let us first formally specify the properties we need \system to guarantee.

\begin{itemize}
	\item \textbf{Safety}
	\begin{itemize}
	\item \textit{Consistency}: If two correct validators $p$ and $q$ commit blocks $b$ and $b'$ respectively to the same slot, then $b$=$b'$.
	\end{itemize}
	\item \textbf{Liveness}
	\begin{itemize}
	\item \textit{Infinite log growth}: Eventually all correct validators commit a value (a block or the special $\bot$ value) for any slot.
	\item \textit{Eventual progress}: After GST the log grows with non $\bot$ values.
	\end{itemize}
\end{itemize}

We argue that the implementation in \Cref{alg:sb-impl-pt1} satisfies the properties of \system.

\subsubsection{Consistency}
\label{sec:consistency}
We prove consistency by contradiction%
\footnote{Slotless blocks inherit consistency from the blocks with slot numbers by which they are committed.}.
Let us assume that validators $p$ and $q$ commit blocks $p$ and $q$ at slot $sn$ with $b\neq b'$.
There exist six potential ways in which the blocks have been committed.
\begin{enumerate}
	\item Both $p$ and $q$ \primitivesmall-delivered $b$ and $b'$ with $sn$.
	\item One of $p$, $q$ \primitivesmall-delivered the corresponding block and the other decided via the fallback protocol.
	\item Both $p$, $q$ decided via the fallback protocol.
	\item One of $p$ and $q$ \primitivesmall-delivered $b$ and the other pulled $b'$ from another validator.
	\item One of $p$ and $q$ decided $b$ via the fallback protocol and the other pulled $b'$ form another validator.
	\item Both $p$, $q$ pulled $b$ and $b'$ from other validators.
\end{enumerate}

In the first case, $p$ and $q$ \primitivesmall-delivered $(id, sn, b$) and $(id', sn, b')$ respectively.
Then, from \primitive unique sequencing property ($\primitiveproperty4$), $b=b'$, a contradiction.

In the second case, let us assume without loss of generality that $p$ \primitivesmall-delivered $id, sn, b$ and $q$ decided $sn,b'$.
Therefore, within \primitive, $p$ received $2f+1$ \ready messages, $f+1$ out of which from a set of correct validators $S_1$.
Let $S_2$ be the set of validators who invoked the \bbcaabort\ events leading to $q$ deciding.
$S_1, S_2$ intersect in at least one correct validator; let's name the correct validator $p_c$.
By belonging in set $S_1$, $p_c$ sent a \ready message for block $b$,
and as argued above already, $b$ is unique with such a ready message for $sn$.
Hence, $q$ must decide on $b$, because $S_2$ includes the ready-certificate $C_{R}$ by $p_c$.
A contradiction.

In the third case, let us assume that $p$ and $q$ decide  $sn, b$ and $sn, b'$ respectively.
By the \vbc safety, no two correct processes decide differently.
Hence, $b=b'$. A contradiction. 

In the fourth case, let's assume without loss of generality that $q$ pulled $b'$ form another validator.
This means that $b'$ was referenced in $f+1$ blocks the $q$ \primitivesmall-delivered such each of them was \primitivesmall-cast by a different validator.
Among the $f+1$ validators at least one is correct. Let's name the correct validator $p_c$.
Therefore, $p_c$ \primitivesmall-delivered $b'$ with $sn$. 
A contradiction follows from the arguments for case (1) for $p$ and $p_c$.
Similarly, case (5), assuming that $p$ decided via the fallback consensus, reduces to case (2) for $p$ and $p_c$.
Finally, in case (6), both blocks $b$ and $b'$ are referenced by $f+1$ blocks.
Therefore, there exist correct validators $p_c$ and $q_c$ who \primitivesmall-delivered $b$ and $b'$ and the contradiction follows from case (1).

\subsubsection{Infinite log growth}
\label{sec:growth}
\lemma{\textit{Totality}: If a correct validator \primitivesmall-delivers a block $(id,sn,b)$ or decides $sn$ for $b$ via the fallback consensus protocol, then every correct validator either \primitivesmall-delivers $(id,sn,b)$ or decides $b$ for slot number $sn$ via the fallback consensus protocol.}
\label{lemma:totality}
\proof{
	If the \primitive sender $p_s$ for instance $id$ is correct, by \primitive integrity, $p_s$ has invoked \primitivesmall-cast for $(id,sn,b)$ and, by \primitive validity, eventually every correct validator delivers $(id,sn,b)$.
	We examine the case where \primitive sender is not correct.
	There are two possible cases; either $sn=\emptyslot$ or $sn\neq\emptyslot$.
	In the first case, $b$ is delivered in the second phase of \primitive, i.e. via a validated reliable broadcast instance which guarantees totality.
	In the second case, if $f+1$ correct validators \primitivesmall-deliver $(id,sn,b)$, then every correct validator \primitivesmall-delivers $(id,sn,b)$ via the \primitive checkpointing mechanism.
	Otherwise, for the \primitive-ledger protocol, $sn$ will cause a timeout for at least $f+1$ correct validators. 
	This is because, by \primitive consistency property, no correct validator will deliver any block other than $b$ for $sn$.
	Therefore, at least $f+1$ correct validators will \primitivesmall-cast a block with the \finalize flag which, by \primitive validity, will be \primitivesmall-delivered by all correct validators.
	Therefore, all correct validators to \primitivesmall-cast a block with the \finalize flag and eventually trigger \vbc for $sn$.
	By \vbc liveness and following the argumentation of \Cref{sec:consistency}, all correct validators will eventually decide $b$ for $sn$.	 
	Finally, if a correct validator decides $sn$ for $b$ via the fallback consensus protocol, then by the properties of the fallback consesus, all correct validators eventually decide $sn$ for $b$.
}

\lemma{\textit{Causality}: If a correct validator $p$ either \primitivesmall-delivers $(id,sn,b)$ or decides $b$ for slot number $sn$ via the fallback consensus protocol, where $b\neq\bot$, then $p$ eventually finalizes $b$ with $sn$.}
\label{lemma:causality}
\proof{
	 Let us assume by contradiction that there exists a causal predecessor $b'$ of $b$ which $p$ never delivers, and therefore $p$ cannot finalize $b$.
	 In the first case, where $p$ either \primitivesmall-delivers, by \primitive external validity property, the external predicate $P$ is valid for $b$, i.e., $p$ has \primitivesmall-delivered $b'$. A contradiction.
	 In the second case, where $p$ decides $b$ via the fallback finalization protocol, since $b\neq\bot$, there exists at least one non empty ready certificate for $b$.
	 This suggests that $2f+1$ validators sent \echo messages for $b$, and therefore, they had already delivered $b'$.
	 Among them, there exist at least $f+1$ correct. 
	 Therefore, by \Cref{lemma:totality}, all correct validators, including $p$, will eventually deliver $p$.
	 A contradiction.
}

Using \Cref{lemma:totality} and \Cref{lemma:causality} we prove the infinite log property for $\snmin$, the first slot number which has not been committed.
Then by a simple induction argument we can show that the property holds for any slot number.

Let's assume by contradiction that some correct validator $p$ never commits a value in slot $\snmin$.
This will eventually trigger a timeout for slot number $\snmin$ and $p$ will trigger $\bbcaabort(\snmin)$.
Let $t$ be the number of correct validators who \primitivesmall-deliver a block $b$ for slot number $\snmin$.
We examine two mutually exclusive sub-cases.
\begin{enumerate}
	\item $t\ge\ f+1$: At least $f+1$ correct validators \primitivesmall-deliver a block $b$ for slot number $\snmin$ and $b\neq\bot$.
	\item $t<f$: There exist at least $f$ more correct validators whose timer for $\snmin$ expires.
\end{enumerate}

In the first case, since the $f+1$ correct validators are correct, by \Cref{lemma:totality}, all correct validators, including $p$, will either \primitivesmall-deliver $b$ with $sn$ or \vbc-decide $b$ for $sn$ and by \Cref{lemma:causality}, $p$ will finalize slot $sn$ with $b$.
Since, by assumption, all slots before $\snmin$ are committed, then $\snmin$ also gets committed.
 A contradiction.

In the second case $f+1$ correct validators will \primitivesmall-bcast a block with a \finalize flag for $\snmin$.
By \primitive validity, all correct validators will \primitivesmall-deliver the blocks, causing eventually all correct validators to \primitivesmall-bcast a block with a \finalize flag for $\snmin$.
Therefore, all correct validators will \primitivesmall-deliver at least $2f+1$ blocks with a \finalize flag for $\snmin$ carrying a valid ready certificate  and will invoke
$\bcpropose([\snmin, b,\mathcal{C}],\mathcal{Q})$,
where $\mathcal{C}$ be set of the ready certificates of the first $2f+1$ blocks which satisfy the predicate $\mathcal{Q}$ for $\snmin$ and some block $b$, possibly also $\bot$.
By \vbc liveness, all correct validators, including $p$, receive $\vbtobdecide([\snmin, b,\mathcal{C}])$.
By \Cref{lemma:causality}, $p$ finalizes $b$ with $\snmin$.
Since $p$ has committed all slots before $\snmin$, it also commits slot $\snmin$.
A contradiction.

\subsubsection{Eventual progress}
\label{sec:growth}
If a correct validator $p$ invokes \primitive for some block $b$ which includes a set of causal references to a set of blocks $B$, such that $p$ has either \primitivesmall-delivered or \vbc-decided for each block in $B$, then, by \Cref{lemma:totality}, all correct validators will \primitivesmall-deliver
	of \vbc-decide for all blocks in $B$.
Therefore, the external validity predicate for \primitive will be satisfied for all correct validators.
Moreover, by the weak sequencing validity property of \primitive ($\primitiveproperty6$), after GST, if some correct validator $p$ invokes \bbcacast\ for some slot number $sn$, all correct validators will \bbcadeliver\ $sn$ with some block $b\neq\bot$, and hence add it to the log, if the following conditions hold: 
\begin{enumerate}
	\item No correct validator invokes \bbcaabort for $sn$.
	\item There exists no other validator who invokes \primitive for $sn$.
\end{enumerate}

The first condition is satisfied, since we are after GST and, therefore no validator will timeout.
The second condition depends on the ticket master implementation.
As we cover in \Cref{sec:ticketmaster}, the ticketmaster functionality ensures that all correct validators are assigned slot numbers and, moreover, slot numbers are assigned exclusively.
The round robin strategy satisfies this by assigning slot numbers to validators in a deterministic way.
In the strategies where the ticket master is a server, we achieve this by rotating the ticket master among validators, so that eventually some correct validator assumes the role.

\section{Conclusion}

\system is a Byzantine log replication technology enabling parallel broadcasts, where each broadcast is sequenced independently and can finalize instantaneously.
While a gap in the sequence numbers will delay the commit of higher sequence numbers, the log continues to be filled, hence the resources of the system are continuously utilized.
Moreover, consensus decisions about gaps in sequence numbers are likewise embedded within broadcasts accumulating is a DAG structure, hence have throughput utility as well.

\system is built around a novel consistent broadcast variant, \primitive,
that attempts to add unique sequence numbers to messages,
while maintaining the safety and liveness guarantees of Byzantine consistent broadcast.
Using \primitive to build a DAG of sequenced messages allows us to commit to a total order of messages extremely fast in the common case.
At the same time, using the DAG to derive consensus under contention, allows our system to maintain the high throughput scalability of state-of-art DAG protocols.


\end{document}